\documentclass[aps,prd,twocolumn,superscriptaddress,footinbib,notitlepage]{revtex4-1}

\usepackage{aas_macros}
\usepackage[utf8]{inputenc}
\usepackage[T1]{fontenc}
\usepackage[english,francais]{babel}
\usepackage{mathrsfs}
\usepackage{amsmath}
\usepackage{amssymb}
\usepackage{tipa}
\usepackage{ntheorem}
\usepackage{braket}
\usepackage{graphicx}
\usepackage{bm}
\usepackage{wasysym}
\usepackage{enumitem}
\usepackage{stmaryrd}
\usepackage[breaklinks=true,colorlinks=true]{hyperref}
\usepackage{tikz}
\usetikzlibrary[patterns]
\usetikzlibrary{shapes}

\theoremseparator{.}
\newtheorem{theorem}{Theorem}

\makeatletter
\newskip\@bigflushglue \@bigflushglue = -100pt plus 1fil

\def\bigcentering{\let\\\@centercr\rightskip\@bigflushglue
\leftskip\@bigflushglue
\parindent\z@\parfillskip\z@skip}

\makeatother

\allowdisplaybreaks[1]

\newcommand{\dd}{\mathrm{d}}

\begin{document}

\title{General expansion of time transfer functions in optical spacetime}
\author{Adrien Bourgoin}
\affiliation{University of Bologna, via fontanelle 40, Forl\`i,  Italy}
\email{adrien.bourgoin@unibo.it}


\begin{abstract}
  When dealing with highly accurate modeling of time and frequency transfers into arbitrarily moving dielectrics medium, it may be convenient to work with Gordon's optical spacetime metric rather than the usual physical spacetime metric. Additionally, an accurate modeling of the geodesic evolution of observable quantities (e.g., the range and the Doppler) requires us to know the reception or the emission time transfer functions. In the physical spacetime, these functions can be derived to any post-Minkowskian orders through a recursive procedure. In this work, we show that the time transfer functions can be determined to any order in Gordon's optical spacetime as well. The exact integral forms of the gravitational, the refractive, and the coupling contributions are recursively derived. The expression of the time transfer function is given within the postlinear approximation assuming a stationary optical spacetime covered with geocentric celestial reference system coordinates. The light-dragging effect due to the steady rotation of the neutral atmosphere of the Earth is found to be at the threshold of visibility in many experiments involving accurate modeling of the time and frequency transfers.
\end{abstract}

\maketitle

\section{Introduction}

In geometrical optics, the concept of light rays is introduced as curves whose tangents coincide with the direction of propagation of an electromagnetic wave \cite{1975ctf..book.....L}. In this approximation, refraction operates at two different levels. First, it causes the phase velocity of the electromagnetic wave to slow down or speed up while crossing a region of higher or lower refractivity, respectively. Secondly, light rays tend to bend toward regions of higher refractivity. These outcomes produce an excess path delay and a geometric delay in the light time. Depending on the context, these two effects must be either thoroughly modeled or precisely measured while designing highly accurate experiments involving time and frequency transfers in the presence of a refractive medium.

In many fields of astronomy such as planetary physics, astrometry, metrology, geodesy, fundamental physics, or even cosmology, we can think of situations where refractivity plays a significant role in the time and frequency transfers. For instance, we mention that ground-based astrogeodetic techniques operating for the realization of the international terrestrial reference frame (ITRF) are currently limited by errors in modeling the group delay during the signal propagation through the Earth's atmosphere \cite{2010ITN....36....1P,2004GeoRL..3114602M,2002GeoRL..29.1414M,1997JGR...10220489C,doi1010292006JB004834}. We also mention the cases of atmospheric radio occultations \cite{1965Sci...149.1243K,1985AJ.....90.1136L,1987JGR....9214987L,1992AJ....103..967L,2012Icar..221.1020S,2015RaSc...50..712S} and atmospheric stellar occultation experiments \cite{1994A&A...288..985R,2006JGRE..11111S91S}. Indeed, both techniques aim at determining a refractivity profile toward an occulting atmosphere from precise measurements of an \emph{a priori} known frequency (usually given in the frame at rest with the emitter) and from an accurate modeling of the frequency transfer in the presence of the occulting refractive medium. To an even higher degree of accuracy, we can cite experiments involving frequency transfers between distant atomic clocks via ground-ground free-space optical (FSO) link \cite{Djerroud.10,2013NaPho...7..434G,doi.10.1063.1.4963130}, space-ground FSO link \cite{Fujieda_2014,Hachisu.14}, and optical fiber links \cite{2010Sci...329.1630C,Predehl441,2015Metro..52..552G}. Finally, let us emphasize that in the context of cosmology, it has been shown that the accumulated effect of an artificial refractivity over the distance-redshift relation perfectly fits the Hubble curve of type Ia supernovae data in the framework of a nonaccelerating cosmological model \cite{PhysRevD.78.044040,*PhysRevD.79.104007,*PhysRevD.80.044019}. All these examples highlight how important refraction can be in highly accurate experiments involving time and frequency transfers.

In the past, two independent theoretical formalisms have been introduced, namely Gordon's optical metric and the time transfer functions. On one side, Gordon's metric allows one to handle refraction in curved spacetime; on the other side, the time transfer functions formalism handles theoretical problems related to the time and frequency transfers in curved spacetime. In this work, we intend to combine the two formalisms which are discussed in turn in the next paragraphs.

In the early 1920s, Gordon introduced \cite{doi101002andp19233772202} a useful theoretical tool to study light refraction caused by an arbitrarily moving fluid dielectric medium, namely, Gordon's optical metric. In this work, he showed that in the presence of a fluid whose electromagnetic properties are described by a permittivity $\epsilon(x)$ and a permeability $\mu(x)$, any solutions to the macroscopic Maxwell's equations can be looked for indifferently either in the usual physical spacetime fitted with the metric tensor, or in an artificial optical spacetime fitted with Gordon's metric. Conveniently, in the optical spacetime and within the geometric optics approximation, by means of a slightly different set of Maxwell's equations, the electromagnetic properties of the fluid medium are reduced to their vacuum values, that is to say $\epsilon(x)=\epsilon_0$ and $\mu(x)=\mu_0$. In other words, in the physical spacetime, the interaction between the electromagnetic field and the dielectric fluid medium must be carefully modeled, whereas in the optical spacetime this interaction is implicitly accounted for in the vacuum limit of the macroscopic version of Maxwell's equations. Consequently, within the geometric optics approximation, light rays propagate into the dielectrics medium along null geodesic lines of the optical spacetime.

At the same time, theoretical problems dealing with the deflection of light rays or the frequency transfer require us to know the function relating the (coordinate) time transfer to the coordinate time at the reception and to the spatial coordinates of the reception and the emission points. Such a function is called a reception time transfer function. Obliviously, an emission time transfer function can be introduced as well. The formalism which aims at determining the time transfer functions was first introduced by Linet and Teyssandier \cite{2002PhRvD..66b4045L} relying on the theory of the world function developed by Synge \cite{SyngeBookGR}. General expansions of the world function and the time transfer functions were first proposed by Le Poncin-Lafitte \emph{et al.} \cite{2004CQGra..21.4463L}, and then a simplified recursive approach, based on the determination of time delay functions instead of Synge's world function, was presented by Teyssandier and Le Poncin-Lafitte \cite{2008CQGra..25n5020T}. The usefulness of the time transfer function formalism lies in the fact that it spares one the trouble of explicitly solving the null geodesic equation which usually leads to heavy calculations beyond the post-Minkowskian regime (see, e.g., \cite{1993tegp.book.....W,1997JMP....38.2587K,1999PhRvD..60l4002K,2001A&A...370..320B,2002PhRvD..65f4025K} for explicit resolution of the null geodesic equation in the linearized weak field limit and see, e.g., \cite{1983PhRvD..28.3007R,1987KFNT....3....8B,Linet_2013} for resolution in the post-post Minkowskian approximation). Indeed, assuming that the emission and reception points events are linked by a null geodesic path (quasi-Minkowskian path approximation), the time transfer functions formalism achieves a complete resolution of the time and frequency transfers to any post-Minkowskian order by means of an algorithmic resolution method \cite{2008CQGra..25n5020T}. For this reason, this formalism is currently one of the most powerful theoretical tools to derive the time and frequency transfers along null geodesics of the curved physical spacetime.

The scope of this paper is to generalize the formalism of the time transfer functions to optical spacetime. The aim is to provide a recursive method allowing one to solve theoretical problems related to the propagation of light in the presence of an arbitrarily moving refractive medium.

This work is organized as follows. In Sec. \ref{sec:not}, we present the notations and conventions used throughout this paper. Section \ref{sec:geO} is a short reminder about the use of Gordon's metric in relativistic geometrical optics. In this section, we derive the optical counterpart of the scalar Eikonal equation (fundamental equation of geometrical optics) which is at the basis of the demonstration which follows. Section~\ref{sec:TF} is a recall about the time transfer functions formalism. In Sec.~\ref{sec:intTF}, by applying a method initially proposed by Teyssandier and Le Poncin-Lafitte \cite{2008CQGra..25n5020T}, we show that working in optical spacetime induces the fact that the time transfer functions can be decomposed into three components, that we call the gravitational, the refractive, and the coupling time transfer functions. In Sec. \ref{sec:exp}, we present the general expansion of the three contributions. In Sec. \ref{sec:app}, we illustrate the method by computing the time transfer function of an optical spacetime describing Earth's rotating atmosphere in the geocentric celestial reference system (GCRS) within the postlinear approximation. Finally, we discuss the importance of taking into account the light-dragging effect in the future generation of data reduction software.

\section{Notations and conventions}
\label{sec:not}

In this work, the metric of spacetime is denoted by $g$ and its signature is $(+,-,-,-)$. The optical metric (also called Gordon's metric) is denoted by $\bar g$.

We suppose that spacetime is covered with some global coordinate system $(x^\mu)=(x^0,x^i)$. We put $x^0=ct$ with $c$ being the speed of light in a vacuum and $t$ being the coordinate time. Greek indices run from 0 to 3 and Latin indices run from 1 to 3.

Straight bold letters (e.g., $\mathbf x$) and italic bold letters (e.g., $\bm x$) denote 3-vectors and 4-vectors, respectively. The 3-vector $\mathbf x$ can also be characterized by an ordered triple of coordinate values $x^i$. Similarly, the 4-vector $\bm x$ can be characterized by an ordered quadruple of coordinate values $x^\mu$. The components of the 4-vector $\bm x$ can be denoted abstractly by $\bm x=(x^0,\mathbf{x})$. When the 4-vector is a separation vector between the origin of the coordinate system and a point event $x$, we make no distinction between the point event and the separation vector. Thus, we associate the point event $x$ with the components $x=(x^0,\mathbf{x})$.

Einstein's summation convention on repeated indices is used for expressions like $a^ib^i$ as well as for expressions like $A^\mu B_\mu$. The ordinary Euclidean norm of $\mathbf{x}$ is denoted as $|\mathbf x|$ and is defined as $|\mathbf{x}|=(\delta_{ij}x^ix^j)^{1/2}$ where $\delta_{ij}$ is the Kronecker delta. The maximum absolute value of the component $A_{\mu\nu}$ is denoted as $|A_{\mu\nu}|_{\mathrm{max}}$. The 3-dimensional antisymmetric Levi-Civita tensor is denoted as $e^{ijk}$.

For the sake of legibility, we employ $(f)_{x}$ or $[f]_{x}$ instead of $f(x)$ whenever necessary. When a quantity $f(x)$ is to be evaluated at two point events $x_A$ and $x_B$, we employ $(f)_{A/B}$ to denote $f(x_A)$ and $f(x_B)$, respectively. The partial differentiation with respect to coordinates $x^\mu$ is denoted $\partial_\mu$. The physical and the optical covariant differentiations with respect to $x^\mu$ are denoted as $\nabla_\mu$ and $\bar\nabla_\mu$, respectively. Given a scalar function $f(x)$, we have the relation $\bar\nabla_\mu f=\nabla_\mu f=\partial_\mu f$.

Throughout the paper, we assume the presence of an arbitrarily moving fluid dielectric medium filling a finite domain $\mathcal{D}$ of spacetime. We call $\bm w(x)$ the unit 4-velocity vector of a point event $x$ belonging to a fluid element of the optical medium. The expression of $\bm w(x)$ is given by
\begin{equation}
  \bm w(x)\equiv\frac{\dd\bm x}{\dd s}
  \label{eq:4velel}
\end{equation}
where the spacetime interval $\dd s$ is defined by
\begin{equation}
  \dd s^2=g_{\mu\nu}(x)\dd x^{\mu} \dd x^{\nu}\text{.}
  \label{eq:ds}
\end{equation}

We call $\xi^i(x)$ the coordinate 3-velocity vector of the point event $x$ belonging to a fluid element of the optical medium. Its expression is given by
\begin{equation}
  \xi^i(x)\equiv\frac{w^i}{w^0}=\frac{1}{c}\frac{\dd x^i}{\dd t}\text{.}
  \label{eq:3velel}
\end{equation}

Finally, $G$ is the Newtonian gravitational constant.

\section{Relativistic Geometrical optics}
\label{sec:geO}

We assume the presence of a fluid optical medium filling $\mathcal{D}$. Additionally, we consider for simplicity that the fluid's electromagnetic properties are linear, isotropic, nondispersive and can be summarized by two scalar functions, namely the permittivity $\epsilon(x)$ and the permeability $\mu(x)$. These two quantities completely determine the refractive properties of the optical medium through the following relationship:
\begin{equation}
  n(x)\equiv c\sqrt{\epsilon(x)\mu(x)}
  \label{eq:n}
\end{equation}
where $n$ is the index of refraction of the medium.

When $x\notin\mathcal{D}$, the permittivity and the permeability reduce to their vacuum values $\epsilon(x)=\epsilon_0$ and $\mu(x)=\mu_0$, respectively. Thus, considering that $c\equiv(\epsilon_0\mu_0)^{-1/2}$, the index of refraction becomes $n(x)=1$. By subtracting its vacuum value from the index of refraction, we obtain the refractivity
\begin{equation}
  N(x)\equiv n(x)-1\text{,}
  \label{eq:N}
\end{equation}
which is obviously null in a vacuum.

In the physical spacetime, the evolution of an electromagnetic phenomenon occurring in the presence of an optical medium is usually described by the macroscopic version of Maxwell's equations. These equations are separated into two distinct sets involving a covariant antisymmetric tensor $F_{\mu\nu}$ called the electromagnetic field tensor (or Faraday tensor), and a contravariant antisymmetric tensor $B^{\mu\nu}$ called the electromagnetic field excitation tensor (or Maxwell tensor), respectively. The macroscopic version of Maxwell's equations are given by \cite{1960ecm..book.....L,doi101002andp200810313}
\begin{subequations}\label{eq:Max}
\begin{align}
  \partial_{[\sigma} F_{\mu\nu]}&=0\text{,}\label{eq:Max1}\\
  \nabla_\mu B^{\mu\nu}&=j^\nu\label{eq:Max2}
\end{align}
\end{subequations}
where $\bm j(x)$ is a 4-vector denoting the free charge density current. The square brackets denote the complete antisymmetrization of the enclosed indices.

The first equation \eqref{eq:Max1} allows one to postulate the existence of a covector field $A_\mu(x)$, such that the electromagnetic field tensor $F_{\mu\nu}$ can be locally written as the rotational of the covector field, that is 
\begin{equation}
  F_{\mu\nu}=\mathrm{Re}\big\{\partial_{\mu}A_\nu-\partial_{\nu}A_\mu\big\}\text{.}
  \label{eq:F}
\end{equation}

The second equation \eqref{eq:Max2} cannot be used alone to fully determine the six independent components of the electromagnetic field excitation tensor $B^{\mu\nu}$. In addition, it does not provide a way to determine the components of the electromagnetic field tensor $F_{\mu\nu}$ which yet governs the motion of particles through the Lorentz force. Therefore, Maxwell's equations must be supplemented with constitutive relations.

For an arbitrarily moving medium of permittivity $\epsilon(x)$ and permeability $\mu(x)$ the covariant constitutive relationships are given by \cite{1960ecm..book.....L}
\begin{subequations}\label{eq:HFctv}
\begin{align}
  B^{\mu\nu}w_\nu&=\epsilon c^2F^{\mu\nu}w_\nu\text{,}\label{eq:Hctv}\\
  \mu B_{[\mu\nu}w_{\sigma]}&=F_{[\mu\nu}w_{\sigma]}\text{.}\label{eq:Fctv}
\end{align}
\end{subequations}

Equations. \eqref{eq:HFctv} can be written as a single relationship involving $B^{\mu\nu}$, $F_{\mu\nu}$, and $\bm w(x)$. Indeed, as initially shown by Gordon \cite{doi101002andp19233772202}, when dealing with problems of electromagnetic waves propagating into dielectrics, it is convenient to introduce an optical spacetime in which refractivity is considered as a spacetime curvature. Gordon's metric (or optical metric) is defined by
\begin{subequations}\label{eq:Gor}
\begin{equation}
  \bar g_{\mu\nu}\equiv g_{\mu\nu}+\gamma_{\mu\nu}\text{,} \qquad \gamma_{\mu\nu}=-\left(1-\frac{1}{n^2}\right)w_\mu w_\nu\text{,}\label{eq:Gorcov}
\end{equation}
with inverse
\begin{equation}
  \bar g^{\mu\nu}\equiv g^{\mu\nu}+\kappa^{\mu\nu}\text{,} \qquad \kappa^{\mu\nu}=(n^2-1)w^\mu w^\nu\text{.}\label{eq:Gorcon}
\end{equation}
\end{subequations}

Making use of Eq. \eqref{eq:Gorcon}, one can see that Eqs. \eqref{eq:HFctv} are summarized within the single following relation \cite{doi101002andp19233772202}:
\begin{equation}
  \mu B^{\mu\nu}=\bar F^{\mu\nu}\text{,}
  \label{eq:ctv}
\end{equation}
where the optical metric has been used to raise covariant indices of $F_{\alpha\beta}$, that is
\begin{equation}
  \bar F^{\mu\nu}\equiv\bar g^{\mu\alpha}\bar g^{\nu\beta} F_{\alpha\beta}\text{.}
  \label{eq:bF}
\end{equation}

It is now possible to express Maxwell's equation in the optical spacetime. Because the covariant components of the electromagnetic field tensor are equivalents in both spacetimes \footnote{This can be easily shown from Eq. \eqref{eq:bF} using the inverse conditions $\bar g_{\mu\sigma}\bar g^{\sigma\nu}=\delta_\mu^\nu$.}, the first pair of Maxwell's equations \eqref{eq:Max1} remains unchanged. The optical form of the second pair \eqref{eq:Max2} is obtained after substituting for $B^{\mu\nu}$ from Eq.~\eqref{eq:ctv} while introducing the optical covariant derivative \cite{PhysRevD.78.044040,PhysRevD.79.104007}. After a little algebra, we find
\begin{equation}
  \bar{\nabla}_\mu\left(\sqrt{\frac{\epsilon}{\mu}}\,\bar F^{\mu\nu}\right)=\sqrt{\epsilon\mu}\,j^{\nu}\text{.}
  \label{eq:bMax}
\end{equation}

Equation \eqref{eq:bMax} is perfectly equivalent to Eq. \eqref{eq:Max2} equipped with the constitutive relations \eqref{eq:HFctv}. While working in the optical spacetime, Eq. \eqref{eq:bMax} allows one to find $\bar F^{\mu\nu}$ and Eq.~\eqref{eq:bF} allows one to express the components of the electromagnetic field tensor in the physical spacetime. Hereafter, we work in the optical spacetime where the light propagation into the dielectric medium is simply given by the vacuum limit of the macroscopic version of Maxwell's equations (no free density current, i.e., $j^\nu=0$). 

In this work, we consider geometrical optics approximation, so we assume that the 4-potential covector $A_\mu(x)$ of a traveling quasimonochromatic wave possesses an expansion of the form \cite{1975ctf..book.....L}
\begin{equation}
  A_\mu=\big[a_\mu+\mathcal{O}(\omega^{-1})\big]e^{i\omega\mathscr{S}}\text{.}
  \label{eq:pot}
\end{equation}
Here $\mathscr{S}(x)$ is the usual eikonal function which determines the surfaces of the constant phase for the wave, $a_\mu(x)$ is the complex covector amplitude varying slowly in comparison to $\mathscr{S}(x)$, and $\omega$ is a bookkeeping parameter that we take to be high during our manipulations \cite{gravitationBook}.

Then substituting for $A_\mu$ from Eq. \eqref{eq:pot} into \eqref{eq:F} allows one to infer
\begin{equation}
  F_{\mu\nu}=\mathrm{Re}\Big\{\big[i\omega f_{\mu\nu}+\mathcal{O}\left(\omega^0\right)\big]e^{i\omega\mathscr{S}}\Big\}\text{,}
  \label{eq:Fopt}  
\end{equation}
where $f_{\mu\nu}(x)$ represent the coordinates of the electromagnetic field tensor amplitude, that is  
\begin{equation}
  f_{\mu\nu}=k_\mu a_\nu-k_\nu a_\mu\text{,}
  \label{eq:A}
\end{equation}
with $k_\mu$ being the wave covector defined by
\begin{equation}
  k_\mu\equiv\partial_\mu\mathscr{S}\text{.}
  \label{eq:kdef}
\end{equation}

We can introduce the contravariant optical wave vector such that
\begin{equation}
  \bar k^\mu\equiv\bar g^{\mu\nu}k_\nu
  \label{eq:bkdef}
\end{equation}
where the low index has been raised with the help of the optical spacetime metric. We can directly check from the inverse conditions $\bar g_{\mu\sigma}\bar g^{\sigma\nu}=\delta_\mu^\nu$ that the covariant coordinates of the wave vector are identical in physical and optical spacetimes; that is to say
\begin{equation}
  \bar k_\mu =k_\mu\text{.}
  \label{eq:kbk}
\end{equation}

Assuming that the 4-potential fulfills the Lorentz gauge in the optical spacetime, that is
\begin{equation}
  \bar\nabla_\mu \bar A^\mu=0
  \label{eq:bLG}
\end{equation} 
where we introduced $\bar A^\mu\equiv\bar g^{\mu\nu}A_\nu$, and we find
\begin{equation}
  \bar g^{\mu\nu}k_\mu a_\nu=0
  \label{eq:bLGopt}
\end{equation}
within the geometrical optics approximation. This relationship states the orthogonality between the optical wave vector $\bar{\bm k}$ and the wave covector amplitude $a_\mu$.

Finally, the fundamental equations of geometrical optics can be derived from the vacuum limit of the optical version of Maxwell's equations. We first determine the optical electromagnetic field tensor by making use of Eqs.~\eqref{eq:bF} and \eqref{eq:Fopt}. Then by taking the covariant derivative of $\bar F^{\mu\nu}$, we find
\begin{equation}
  \bar\nabla_\mu\bar F^{\mu\nu}=-\mathrm{Re}\Big\{\big[\omega^2k_\mu\bar f^{\mu\nu}+\mathcal{O}(\omega)\big]e^{i\omega\mathscr{S}}\Big\}
  \label{eq:CDFopt}  
\end{equation}
where we introduced $\bar f^{\mu\nu}\equiv\bar g^{\mu\alpha}\bar g^{\nu\beta}f_{\alpha\beta}$. By substituting this result into the vacuum limit of Eq. \eqref{eq:bMax} and by restricting ourselves to the geometrical optics order, we deduce
\begin{equation}
  \mathrm{Re}\big\{k_\mu\bar f^{\mu\nu}\big\}=0\text{.}
  \label{eq:beikA}
\end{equation}

Then substituting for $\bar f^{\mu\nu}$ from Eq. \eqref{eq:A} into \eqref{eq:beikA} and considering \eqref{eq:bLGopt}, we finally deduce
\begin{equation}
  \bar g^{\mu\nu}k_\mu k_\nu=0\text{.}
  \label{eq:beik}
\end{equation}
This is the fundamental equation of geometrical optics expressed in optical spacetime. After inserting Eq. \eqref{eq:kdef}, we infer that the phase $\mathscr{S}(x)$ satisfies the well known scalar Eikonal equation
\begin{equation}
  \bar g^{\mu\nu}\partial_\mu\mathscr{S}\partial_\nu\mathscr{S}=0\text{.}
  \label{eq:beikS}
\end{equation}

We close this section by showing that $\bar{\bm k}$ is a null vector satisfying the geodesic equation for the optical metric. From Eqs. \eqref{eq:beik} and \eqref{eq:kbk}, we easily infer
\begin{equation}
  \bar g_{\mu\nu}\bar k^\mu \bar k^\nu=0\text{.}
  \label{eq:beikcont}
\end{equation}
This relation shows that $\bar{\bm k}$ is indeed isotropic for the optical metric $\bar g_{\mu\nu}$. Then we differentiate Eq. \eqref{eq:beik} with respect to $x^\sigma$. Considering the symmetry of the components of the optical metric together with Eq.~\eqref{eq:bkdef}, it becomes
\begin{equation}
  \bar k^\nu(\bar\nabla_\sigma k_\nu)=0\text{.}
\end{equation}
Making use of the definition \eqref{eq:kdef}, we infer $\bar\nabla_\sigma k_\nu=\bar\nabla_\nu k_\sigma$. Finally, considering Eqs. \eqref{eq:bkdef} and \eqref{eq:kbk}, we deduce
\begin{equation}
  (\bar k^\nu\bar\nabla_\nu) \bar k^\sigma=0\text{,}
  \label{eq:geod}
\end{equation}
which states [together with Eq. \eqref{eq:beikcont}] that curves admitting $\bar{\bm k}$ as a tangent vector are null geodesic lines of the optical metric. In that respect, a null line which is the solution of Eq. \eqref{eq:geod} can be interpreted as a ray of light whose tangent at any point $x$ is orthogonal to the surface of the constant phase $\mathscr{S}(x)$ \cite{SyngeBookGR}.

\section{Time transfer functions formalism}
\label{sec:TF}

Let us consider a light ray $\Gamma$ propagating in a region of spacetime covered with some coordinate system $(x^\mu)$. Let $(ct_A,\mathbf{x}_A)$ be the components of the point event $x_A$. We introduce $\mathcal{C}_A$, the curve of parametric equations $x=x_A(\tau)$ with $\tau$ being a parametrization of $\mathcal{C}_A$. Let us suppose that the coordinate system is chosen such that $\mathcal{C}_A$ is a timelike worldline for any $x_A$, which means that $\partial/\partial x^0$ is a timelike vector field, that is to say $g_{00}>0$ everywhere. Let $x_A$ be the point event where $\Gamma$ is emitted and let $x_B$ be the point event of components $(ct_B,\mathbf{x}_B)$ where it is observed. The quantity $t_B-t_A$ is the (coordinate) travel time of the light ray connecting the emission point event $x_A$ and the reception point event $x_B$. This quantity allows us to introduce the time transfer functions $\mathcal{T}_{r,\Gamma}$ and $\mathcal{T}_{e,\Gamma}$ \cite{2004CQGra..21.4463L} as 
\begin{equation}
  t_B-t_A\equiv\mathcal{T}_{r,\Gamma}(\mathbf x_A,t_B,\mathbf x_B)\equiv\mathcal{T}_{e,\Gamma}(t_A,\mathbf x_A,\mathbf x_B)\text{.}
  \label{eq:TTFdefGen}
\end{equation}
We call $\mathcal{T}_{r,\Gamma}$ the reception time transfer function and $\mathcal{T}_{e,\Gamma}$ the emission time transfer function associated with $\Gamma$.

As shown in \cite{PhysRevD.93.044028}, given a point event $x_B$ and a spatial position $\mathbf x_A$, $\Gamma$ is not unique in general. Thus, let $\{\Gamma^{[\sigma]}(\mathbf x_A,x_B)\}_{\sigma\in\mathbb{N}}$ be a family of light rays intersecting $x_B$ and flowing from the different point events
\begin{equation}
  x_A^{[\sigma]}\in\mathcal{C}_A\text{,} \qquad x_A^{[\sigma]}=(ct_A^{[\sigma]},\mathbf x_A)\text{.}
\end{equation}
For each $\Gamma^{[\sigma]}$, there exists a reception time transfer function, denoted by $\mathcal{T}_{r,\Gamma^{[\sigma]}}(\mathbf x_A,t_B,\mathbf{x}_B)$, such that
\begin{equation}
  t_B-t_A^{[\sigma]}=\mathcal{T}_{r,\Gamma^{[\sigma]}}(\mathbf x_A,t_B,\mathbf{x}_B)
\end{equation}
(the same reasoning works for the emission time transfer function as well).

This fact shows that, in general, we cannot expect to find a unique reception (or emission) time transfer function. However, for a very particular type of null geodesics, referred to as quasi-Minkowskians \cite{2012CQGra..29x5010T,PhysRevD.93.044028}, it has been shown that the reception (or the emission) time transfer function, if it exists, can be uniquely determined \cite{2008CQGra..25n5020T}.


Henceforth, we assume that $\Gamma$ is a quasi-Minkowskian light ray so that the corresponding time transfer functions are indeed unique. In agreement with this assumption, we suppose that the past null cone at $x_B$ intersects $\mathcal{C}_A$ at one and only one point $x_A$. Therefore, we can rewrite Eq.~\eqref{eq:TTFdefGen} as
\begin{equation}
  t_B-t_A\equiv\mathcal{T}_r(\mathbf x_A,t_B,\mathbf x_B)\equiv\mathcal{T}_e(t_A,\mathbf x_A,\mathbf x_B)\text{.}
  \label{eq:TTFdef}
\end{equation}

Hereafter, in order to shorten future notations, we introduce the reception and the emission range transfer functions being defined by
\begin{subequations}\label{eq:RTFdef}
\begin{equation}
  \mathcal{R}_r(\mathbf x_A,x_B)\equiv c\mathcal{T}_r(\mathbf x_A,t_B,\mathbf x_B)\text{,}
  \label{eq:RTFrdef}
\end{equation}
and
\begin{equation}
  \mathcal{R}_e(x_A,\mathbf x_B)\equiv c\mathcal{T}_e(t_A,\mathbf x_A,\mathbf x_B)\text{.}
  \label{eq:RTFedef}
\end{equation}
\end{subequations}

An important theorem (cf. Theorem 1 of \cite{2004CQGra..21.4463L}) states that the covariant coordinates of the tangent vector are totally known as soon as one of the time transfer functions (or equivalently, one of the range transfer functions) is explicitly determined. Therefore, if we define
\begin{equation}
  (l_i)_{A/B}\equiv \left(\frac{k_i}{k_0}\right)_{A/B}\text{,}
  \label{eq:l}
\end{equation}
we have the following relationships:
\begin{subequations}\label{eq:kTFdef}
\begin{align}
  (l_i)_A&\equiv\frac{\partial\mathcal{R}_r}{\partial x_A^i}=\frac{\partial\mathcal{R}_e}{\partial x_A^i}\left(1+\frac{\partial\mathcal{R}_e}{\partial x_A^0}\right)^{-1}\text{,}\label{eq:kTFA}\\
  (l_i)_B&\equiv-\frac{\partial\mathcal{R}_r}{\partial x_B^i}\left(1-\frac{\partial\mathcal{R}_r}{\partial x_B^0}\right)^{-1}=-\frac{\partial\mathcal{R}_e}{\partial x_B^i}\text{,}\label{eq:kTFB}
\end{align}
and
\begin{equation}
  \frac{(k_0)_B}{(k_0)_A}\equiv1-\frac{\partial\mathcal{R}_r}{\partial x_B^0}=\left(1+\frac{\partial\mathcal{R}_e}{\partial x_A^0}\right)^{-1}\text{.}
  \label{eq:kTF0}
\end{equation}
\end{subequations}

Consequently, Eqs. \eqref{eq:kTFdef} completely solve theoretical problems related to frequency transfer. Indeed, it is well known that the instantaneous expression of the Doppler shift along the null-geodesic path between the emitter and the receiver can be expressed as \cite{SyngeBookGR}
\begin{equation}
  \frac{\nu_B}{\nu_A}\equiv\frac{(u^{\mu}k_\mu)_B}{(u^{\mu}k_\mu)_A}=\frac{(u^0k_0)_B}{(u^0k_0)_A}\frac{(1+\beta^il_i)_B}{(1+\beta^il_i)_A}
  \label{eq:dop}
\end{equation}\\
where $(\bm u)_{A/B}$ is the emitter/receiver's unit 4-velocity vectors being defined as
\begin{equation}
  (\bm u)_{A/B}\equiv\left(\frac{\dd\bm x}{\dd s}\right)_{A/B}\text{,}
  \label{eq:4velAB}
\end{equation}
with $\dd s$ introduced in Eq. \eqref{eq:ds}.

By definition, the 4-velocities satisfy the unity condition $(g_{\mu\nu}u^\mu u^\nu)_{A/B}=1$, which implies
\begin{equation}
  (u^0)_{A/B}=(g_{00}+2g_{0i}\beta^i+g_{ij}\beta^i\beta^j)^{-1/2}_{A/B}\text{.}
  \label{eq:4vel}
\end{equation}
The quantities $(\beta^i)_{A/B}$ in Eq. \eqref{eq:dop} represent the coordinates of the emitter/receiver's coordinate 3-velocity vectors and are defined such that
\begin{equation}
  (\beta^i)_{A/B}\equiv\left(\frac{u^i}{u^0}\right)_{A/B}=\frac{1}{c}\left(\frac{\dd x^i}{\dd t}\right)_{A/B}\text{.}
  \label{eq:3velAB}
\end{equation}

It is then straightforward to determine the exact expression of the instantaneous Doppler formulation in terms of the range transfer functions \cite{2001A&A...370..320B,2012CQGra..29w5027H,2014PhRvD..89f4045H}. Indeed, after inserting Eqs.~\eqref{eq:kTFdef} and \eqref{eq:4vel} into Eq. \eqref{eq:dop}, we infer
\begin{equation}
  \frac{\nu_B}{\nu_A}=\frac{(u^0)_B}{(u^0)_A}\frac{q_B}{q_A}\text{,}
  \label{eq:dopTF}
\end{equation}
with
\begin{subequations}\label{eq:qTF}
\begin{align}
  q_A&=1+\beta^i_A\frac{\partial\mathcal{R}_r}{\partial x_A^i}=1+\frac{\partial\mathcal{R}_e}{\partial x_A^0}+\beta^i_A\frac{\partial\mathcal{R}_e}{\partial x_A^i}\text{,}\label{eq:qA}\\
  q_B&=1-\frac{\partial\mathcal{R}_r}{\partial x_B^0}-\beta^i_B\frac{\partial\mathcal{R}_r}{\partial x_B^i}=1-\beta^i_B\frac{\partial\mathcal{R}_e}{\partial x_B^i}\text{,}\label{eq:qB}
\end{align}
\end{subequations}
and
\begin{equation}
  \frac{(u^0)_B}{(u^0)_A}=\frac{(g_{00}+2g_{0i}\beta^i+g_{ij}\beta^i\beta^j)^{1/2}_{A}}{(g_{00}+2g_{0i}\beta^i+g_{ij}\beta^i\beta^j)^{1/2}_{B}}\text{.}
  \label{eq:uBuA}
\end{equation}

From the fundamental equation of geometrical optics [see Eq. \eqref{eq:beik}], we know that the covariant coordinates of the 4-wave optical vector at point events $x_A$ or $x_B$ satisfy a relation as follows
\begin{equation}
  (\bar g^{\mu\nu}k_\mu k_\nu)_{A/B}=0\text{.}
  \label{eq:beikAB}
\end{equation}
Then dividing by $[(k_0)_{A/B}]^2$ and making use of Eqs. \eqref{eq:TTFdef}--\eqref{eq:kTFdef}, we infer the following theorem which generalizes Theorem~1 of \cite{2008CQGra..25n5020T} to optical spacetime.

\begin{widetext}
  \begin{theorem}
    Within geometrical optics approximation, the range transfer functions $\mathcal{R}_r$ and $\mathcal{R}_e$ satisfy the following Hamilton-Jacobi-like equations over the optical spacetime, namely:
    \begin{subequations}\label{eq:beikHJ}
    \begin{equation}
      \bar g^{00}(x_B^0-\mathcal{R}_r,\mathbf{x}_A)+2\bar g^{0i}(x_B^0-\mathcal{R}_r,\mathbf{x}_A)\frac{\partial\mathcal{R}_r}{\partial x_A^i}+\bar g^{ij}(x_B^0-\mathcal{R}_r,\mathbf{x}_A)\frac{\partial\mathcal{R}_r}{\partial x_A^i}\frac{\partial\mathcal{R}_r}{\partial x_A^j}=0\text{,}\label{eq:beikHJA}
    \end{equation}
    and
    \begin{equation}
      \bar g^{00}(x_A^0+\mathcal{R}_e,\mathbf{x}_B)-2\bar g^{0i}(x_A^0+\mathcal{R}_e,\mathbf{x}_B)\frac{\partial\mathcal{R}_e}{\partial x_B^i}+\bar g^{ij}(x_A^0+\mathcal{R}_e,\mathbf{x}_B)\frac{\partial\mathcal{R}_e}{\partial x_B^i}\frac{\partial\mathcal{R}_e}{\partial x_B^j}=0\text{,}\label{eq:beikHJB}
    \end{equation}
    \end{subequations}
    respectively.
    \label{th:theo1}
  \end{theorem}
\end{widetext}

This theorem is at the basis of the demonstration for deriving the integral form of the range and then the time transfer functions. Henceforth, in order to avoid repetitions, we pursue the demonstration giving details only for the reception time delay function. However, the same results can be derived for the emission time delay function by applying the exact same reasoning.

\section{Integral form of the time delay functions}
\label{sec:intTF}

Now let us assume that the physical spacetime metric takes the following form
\begin{subequations}\label{eq:gm}
\begin{equation}
  g_{\mu\nu}=\eta_{\mu\nu}+h_{\mu\nu}
  \label{eq:gmcov}
\end{equation}
throughout spacetime, where $\eta_{\mu\nu}$ is the Minkowski metric and $h_{\mu\nu}$ is the gravitational perturbation. In Cartesian coordinates, $\eta_{\mu\nu}=\text{diag}(+1,-1,-1,-1)$. The contravariant components of the physical spacetime metric can be decomposed as
\begin{equation}
  g^{\mu\nu}=\eta^{\mu\nu}+k^{\mu\nu}
  \label{eq:gmcon}
\end{equation}
\end{subequations}
where the components $k^{\mu\nu}$ satisfy
\begin{equation}
  k^{\mu\nu}=-\eta^{\mu\alpha}\eta^{\beta\nu}h_{\alpha\beta}-\eta^{\mu\alpha}h_{\alpha\beta}k^{\beta\nu}\text{.}
  \label{eq:kmcon}
\end{equation}

Therefore, the optical spacetime metric \eqref{eq:Gorcov} can be expressed as
\begin{subequations}\label{eq:Gordec}
\begin{equation}
  \bar g_{\mu\nu}=\eta_{\mu\nu}+H_{\mu\nu}\text{,}
  \label{eq:Gordeccov}
\end{equation}
with the contravariant components
\begin{align}
  \bar g^{\mu\nu}&=\eta^{\mu\nu}+K^{\mu\nu}\text{.}
  \label{eq:Gordeccon}
\end{align}
\end{subequations}
Thus, the optical metric reduces to the sum of the flat Minkowski metric plus a spacetime curvature contribution which is given by
\begin{subequations}\label{eq:Cur}
\begin{equation}
  H_{\mu\nu}=h_{\mu\nu}+\gamma_{\mu\nu}\text{,}\label{eq:Curcov}
\end{equation}
with the contravariant components
\begin{equation}
  K^{\mu\nu}=k^{\mu\nu}+\kappa^{\mu\nu}\text{.}\label{eq:Curcon}
\end{equation}
\end{subequations}

From here we suppose that the curvature contribution is small so that spacetime is mainly flat; that is to say
\begin{equation}
  |h_{\mu\nu}|_{\mathrm{max}}\ll|\eta_{\mu\nu}|_{\mathrm{max}}\text{,} \qquad |\gamma_{\mu\nu}|_{\mathrm{max}}\ll|\eta_{\mu\nu}|_{\mathrm{max}}\text{.}\label{eq:approxPM}
\end{equation}
In other words, we focus on the post-Minkowskian approximation. Under this condition, we ensure that the null geodesic path is quasi-Minkowskian.

The form of the optical metric in Eqs. \eqref{eq:Gordec} implies that the reception and the emission range transfer functions can be looked for according to the following expressions:
\begin{subequations}\label{eq:RTFdec}
\begin{equation}
  \mathcal{R}_r(\mathbf{x}_A,x_B)=|\mathbf{x}_B-\mathbf{x}_A|+\Delta(\mathbf{x}_A,x_B)\text{,}\label{eq:RTFdecR}
\end{equation}
and
\begin{equation}
  \mathcal{R}_e(x_A,\mathbf{x}_B)=|\mathbf{x}_B-\mathbf{x}_A|+\Xi(x_A,\mathbf{x}_B)\text{,}\label{eq:RTFdecE}
\end{equation}
\end{subequations}
respectively. Following \cite{2008CQGra..25n5020T}, we will call $\Delta/c$ the reception time delay function and $\Xi/c$ the emission time delay function \footnote{The reception and the emission time delay functions in \cite{2008CQGra..25n5020T} are denoted $\Delta_r/c$ and $\Delta_e/c$, respectively. Instead, in this work, we use $\Delta/c$ and $\Xi/c$ in order to keep incoming indices notations as clear as possible.}.

Now if we assume that the reception point event $x_B$ is perfectly known, then we can regard its components $t_B$ and $\mathbf x_B$ as fixed parameters. Hence, the reception time delay function becomes a function of the spatial components of the emission point event $x_A$ \footnote{We recall that the time component of $x_A$ is constrained by the fact that the past null cone at $x_B$ intersects the worldline $\mathbf x=\mathbf x_A$ at only one point of coordinates $x_A=(ct_A,\mathbf{x}_A)$. This is true as long as the null geodesic is quasi-Minkowskian.}. Thus, if we now substitute $\mathbf x$ to $\mathbf x_A$, the reception time delay function $\Delta(\mathbf x,x_B)/c$ uniquely defines the point event $x_-(\mathbf x)$ for the given set of spatial components $\mathbf{x}$, that is to say
\begin{equation}
  x_-(\mathbf x)=\big(x_B^0-|\mathbf x_B-\mathbf x|-\Delta(\mathbf{x},x_B),\mathbf{x}\big)\text{.}
  \label{eq:x-}
\end{equation}
Furthermore, assuming that the point event $x_-$ lies in the vicinity of $x_B$, we can determine the spatial variation of the reception time delay function. Indeed, after inserting Eqs.~\eqref{eq:RTFdecR} and \eqref{eq:Gordeccon} into \eqref{eq:beikHJA} taken at $x_-$ instead of $x_A$, we deduce the following relationship \cite{2008CQGra..25n5020T}
\begin{equation}
  -2N^i\partial_i\Delta(\mathbf{x},x_B)=\Omega_-(x_-,x_B)
  \label{eq:beikDP}
\end{equation}
where $\mathbf{N}=(\mathbf{x}_B-\mathbf{x})/|\mathbf{x}_B-\mathbf{x}|$, and
\begin{align}
  \Omega_-(x_-&,x_B)=\big(K^{00}-2K^{0i}N^i+K^{ij}N^iN^j\big)_{x_-}\nonumber\\
  &+2\big(K^{0i}-K^{ij}N^j\big)_{x_-}\partial_i\Delta(\mathbf{x},x_B)\nonumber\\
  &+\big(\eta^{ij}+K^{ij}\big)_{x_-}\partial_i\Delta(\mathbf{x},x_B)\partial_j\Delta(\mathbf{x},x_B)\text{.}\label{eq:Om}
\end{align}

Since $\mathbf x$ is a free variable, we follow \cite{2008CQGra..25n5020T} and choose for convenience to focus on the case where $\mathbf x$ is varying along the straight line segment connecting $\mathbf x_A$ to $\mathbf x_B$, that is to say $\mathbf{x}=\mathbf{z}_-(\lambda)$, where
\begin{equation}
  \mathbf{z}_-(\lambda)=\mathbf{x}_B-\lambda R_{AB}\mathbf{N}_{AB}\text{,} \qquad 0\leqslant\lambda\leqslant 1\text{,}\label{eq:z-i}
\end{equation}
with $R_{AB}=|\mathbf{x}_B-\mathbf{x}_A|$ and $\mathbf N_{AB}=(\mathbf{x}_{B}-\mathbf{x}_{A})/R_{AB}$. In that respect, we also have the relation
\begin{equation}
  \mathbf{N}=\mathbf N_{AB}\text{.}
  \label{eq:NAB}
\end{equation}

We can now determine the integral form of the time delay function by differentiating $\Delta(\mathbf{z}_-(\lambda),x_B)$ with respect to $\lambda$. Using Eq. \eqref{eq:z-i}, we can always write
\begin{equation}
  \frac{\dd}{\dd\lambda}\Delta(\mathbf{z}_-(\lambda),x_B)=-R_{AB}N_{AB}^i\big[\partial_i\Delta\big]_{(\mathbf{z}_-(\lambda),x_B)}
  \label{eq:DTdel}
\end{equation}
where $[\partial_i\Delta]_{(\mathbf{z}_-(\lambda),x_B)}$ denotes the partial derivative of $\Delta(\mathbf{x},x_B)$ with respect to $x^i$ taken at $\mathbf x=\mathbf z_-(\lambda)$. Then after inserting Eqs. \eqref{eq:beikDP} and \eqref{eq:NAB} into \eqref{eq:DTdel}, we infer 
\begin{equation}
  \frac{\dd}{\dd\lambda}\Delta(\mathbf{z}_-(\lambda),x_B)=\frac{R_{AB}}{2}\,\Omega_-(\widetilde{z}_-(\lambda),x_B)
  \label{eq:DTdeleik}
\end{equation}
where the components of the point event $\widetilde{z}_-(\lambda)$ are obtained from Eq. \eqref{eq:x-} which states that $\widetilde{z}_-(\lambda)=x_-(\mathbf z_-(\lambda))$. They are explicitly written later on in Eq.~\eqref{eq:tz-m}.

By fixing the following boundary conditions:
\begin{subequations}\label{eq:bdR}
\begin{align}
  \Delta(\mathbf{z}_-(0),x_B)&=0\text{,}\label{eq:bdR0}\\
  \Delta(\mathbf{z}_-(1),x_B)&=\Delta(\mathbf{x}_A,x_B)\text{,}\label{eq:bdR1}
\end{align}
\end{subequations}
which follow from the requirement that $\Delta(\mathbf{x}_B,x_B)=0$ when $\mathbf{z}_-(0)=\mathbf{x}_B$, we find
\begin{equation}
  \Delta(\mathbf{x}_A,x_B)=\frac{R_{AB}}{2}\int_0^1\Omega_-(\widetilde{z}_-(\lambda),x_B)\,\dd\lambda\text{.}
  \label{eq:intdel}
\end{equation}
Then insertion of Eq. \eqref{eq:Om} allows us to recover the Theorem 2 of \cite{2008CQGra..25n5020T} which would be expressed here in terms of the contravariant components $K^{\mu\nu}$ instead of $k^{\mu\nu}$'s.

In principle, all machinery developed in \cite{2008CQGra..25n5020T} for computing the delay functions could be applied directly using the components $K^{\mu\nu}$. However, such an approach possesses the inconvenience of hiding the role played by the different components $k^{\mu\nu}$ and $\kappa^{\mu\nu}$ during the determination of the total delay functions.

Indeed, according to Eqs. \eqref{eq:Cur} the curvature of the optical spacetime is described simultaneously with the help of the components $h_{\mu\nu}$ and $\gamma_{\mu\nu}$ which might act on different characteristic lengths [e.g., $\gamma_{\mu\nu}(x)=0$ for $x\notin\mathcal{D}$] and might possess completely different orders of magnitude \emph{a priori}. Therefore, in order to disentangle the contribution of each perturbation into the determination of the total delay, we must perform a complete separation between the physical quantities in Eq. \eqref{eq:intdel}.

As might be seen from Eqs. \eqref{eq:Om} and \eqref{eq:Curcon} such a separation can be achieved when the total time delay functions take the following forms
\begin{subequations}\label{eq:del}
\begin{align}
  \Delta(\mathbf{x}_A,x_B)&=\Delta_{\mathrm{g}}(\mathbf{x}_A,x_B)+\Delta_{\mathrm{r}}(\mathbf{x}_A,x_B)\nonumber\\
  &+\Delta_{\mathrm{gr}}(\mathbf{x}_A,x_B)\text{,}\label{eq:delR}
\end{align}
and
\begin{align}
  \Xi(x_A,\mathbf{x}_B)&=\Xi_{\mathrm{g}}(x_A,\mathbf{x}_B)+\Xi_{\mathrm{r}}(x_A,\mathbf{x}_B)\nonumber\\
  &+\Xi_{\mathrm{gr}}(x_A,\mathbf{x}_B)\text{.}
  \label{eq:delE}
\end{align}
\end{subequations}
The subscripts ``$\mathrm{g}$'', ``$\mathrm{r}$'', and ``$\mathrm{gr}$'' refer to the gravitational, the refractive, and the coupling contributions, respectively.

The gravitational and the refractive time delay functions is expected to be driven by gravitational and refractive perturbations, respectively. Instead, the coupling time delay functions is expected to be of the order of the product of both perturbations.

By substituting for $\Delta(\mathbf{x}_A,x_B)$ from Eq. \eqref{eq:delR} into \eqref{eq:intdel} and \eqref{eq:Om}, and then by making use of the contravariant components of the optical and the physical spacetime metrics [see Eq.~\eqref{eq:Curcon}], we deduce the following theorem.

\begin{widetext}
  \begin{theorem}
    In the optical spacetime, the function $\Delta$ introduced in Eq.~\eqref{eq:RTFdecR}, can be decomposed as shown in Eq.~\eqref{eq:delR} where each term in the summation satisfies an integrodifferential equation 
    \begin{subequations}\label{eq:intdeltR}
    \begin{align}
      \Delta_{\mathrm{g}}(\mathbf{x}_A,x_B)&=\frac{R_{AB}}{2}\int_0^1\bigg\{\big(k^{00}-2k^{0i}N^i_{AB}+k^{ij}N^i_{AB}N^j_{AB}\big)_{\widetilde{z}_-(\lambda)}\nonumber\\
      &+2\big(k^{0i}-k^{ij}N^j_{AB}\big)_{\widetilde{z}_-(\lambda)}\left[\frac{\partial\Delta_{\mathrm{g}}}{\partial x^i}\right]_{(\mathbf{z}_-(\lambda),x_B)}+\big(\eta^{ij}+k^{ij}\big)_{\widetilde{z}_-(\lambda)}\left[\frac{\partial\Delta_{\mathrm{g}}}{\partial x^i}\frac{\partial\Delta_{\mathrm{g}}}{\partial x^j}\right]_{(\mathbf{z}_-(\lambda),x_B)}\bigg\}\dd\lambda\text{,}\label{eq:intdeltRg}\\
      \Delta_{\mathrm{r}}(\mathbf{x}_A,x_B)&=\frac{R_{AB}}{2}\int_0^1\bigg\{\big(\kappa^{00}-2\kappa^{0i}N^i_{AB}+\kappa^{ij}N^i_{AB}N^j_{AB}\big)_{\widetilde{z}_-(\lambda)}\nonumber\\
      &+2\big(\kappa^{0i}-\kappa^{ij}N^j_{AB}\big)_{\widetilde{z}_-(\lambda)}\left[\frac{\partial\Delta_{\mathrm{r}}}{\partial x^i}\right]_{(\mathbf{z}_-(\lambda),x_B)}+\big(\eta^{ij}+\kappa^{ij}\big)_{\widetilde{z}_-(\lambda)}\left[\frac{\partial\Delta_{\mathrm{r}}}{\partial x^i}\frac{\partial\Delta_{\mathrm{r}}}{\partial x^j}\right]_{(\mathbf{z}_-(\lambda),x_B)}\bigg\}\dd\lambda\text{,}\label{eq:intdeltRr}
    \end{align}
    and
    \begin{align}
      \Delta_{\mathrm{gr}}(\mathbf{x}_A,x_B)&=\frac{R_{AB}}{2}\int_0^1\bigg\{\big(\eta^{ij}+k^{ij}+\kappa^{ij}\big)_{\widetilde{z}_-(\lambda)}\left[\frac{\partial\Delta_{\mathrm{gr}}}{\partial x^i}\frac{\partial\Delta_{\mathrm{gr}}}{\partial x^j}\right]_{(\mathbf{z}_-(\lambda),x_B)}\nonumber\\
      &+2\big(k^{0i}+\kappa^{0i}-(k^{ij}+\kappa^{ij})N^j_{AB}\big)_{\widetilde{z}_-(\lambda)}\left[\frac{\partial\Delta_{\mathrm{gr}}}{\partial x^i}\right]_{(\mathbf{z}_-(\lambda),x_B)}\nonumber\\
      &+\big(k^{ij}\big)_{\widetilde{z}_-(\lambda)}\left[\frac{\partial\Delta_{\mathrm{r}}}{\partial x^i}\frac{\partial\Delta_{\mathrm{r}}}{\partial x^j}\right]_{(\mathbf{z}_-(\lambda),x_B)}+2\big(k^{0i}-k^{ij}N^j_{AB}\big)_{\widetilde{z}_-(\lambda)}\left[\frac{\partial\Delta_{\mathrm{r}}}{\partial x^i}\right]_{(\mathbf{z}_-(\lambda),x_B)}\nonumber\\
      &+\big(\kappa^{ij}\big)_{\widetilde{z}_-(\lambda)}\left[\frac{\partial\Delta_{\mathrm{g}}}{\partial x^i}\frac{\partial\Delta_{\mathrm{g}}}{\partial x^j}\right]_{(\mathbf{z}_-(\lambda),x_B)}+2\big(\kappa^{0i}-\kappa^{ij}N^j_{AB}\big)_{\widetilde{z}_-(\lambda)}\left[\frac{\partial\Delta_{\mathrm{g}}}{\partial x^i}\right]_{(\mathbf{z}_-(\lambda),x_B)}\nonumber\\
      &+2\big(\eta^{ij}+k^{ij}+\kappa^{ij}\big)_{\widetilde{z}_-(\lambda)}\left[\frac{\partial\Delta_{\mathrm{g}}}{\partial x^i}\frac{\partial\Delta_{\mathrm{r}}}{\partial x^j}+\frac{\partial\Delta_{\mathrm{g}}}{\partial x^i}\frac{\partial\Delta_{\mathrm{gr}}}{\partial x^j}+\frac{\partial\Delta_{\mathrm{r}}}{\partial x^i}\frac{\partial\Delta_{\mathrm{gr}}}{\partial x^j}\right]_{(\mathbf{z}_-(\lambda),x_B)}\bigg\}\dd\lambda\text{.}\label{eq:intdeltRgr}
    \end{align}
    \end{subequations}
    The components of the point event $\widetilde{z}_-(\lambda)$ are given by
    \begin{equation}
      \widetilde{z}_-(\lambda)=\big(x_B^0-\lambda R_{AB}-\Delta(\mathbf z_-(\lambda),x_B),\mathbf z_-(\lambda)\big)\text{,}
      \label{eq:tz-m}
    \end{equation}
    where $\mathbf z_-(\lambda)$ is defined as in Eq. \eqref{eq:z-i}.
    \label{th:theo2}
  \end{theorem}
\end{widetext}

Following the exact same reasoning, we state a similar theorem for the emission time delay function. However, for the emission case, the straight line segment connecting the emitter $x_A$ to the receiver $x_B$ is defined by \footnote{Hereafter, $\mu$ is an affine parameter and does not represent the permeability of the dielectric medium anymore.}
\begin{align}
  \mathbf z_+(\mu)&=\mathbf x_A+\mu R_{AB}\mathbf{N}_{AB}\text{,} & 0\leqslant\mu\leqslant 1\text{.}\label{eq:z+i}
\end{align}
Then from the requirement that $\Xi(x_A,\mathbf{x}_A)=0$ when $\mathbf{z}_+(0)=\mathbf{x}_A$, we can set the following boundary conditions
\begin{subequations}\label{eq:bdE}
\begin{align}
  \Xi(x_A,\mathbf{z}_+(0))&=0\text{,}\label{eq:bdE0}\\
  \Xi(x_A,\mathbf{z}_+(1))&=\Xi(x_A,\mathbf{x}_B)\text{.}\label{eq:bdE1}
\end{align}
\end{subequations}

Hence, the theorem for the emission time delay function $\Xi(x_A,\mathbf x_B)/c$ reads as follows.

\begin{widetext}
  \begin{theorem}
    In the optical spacetime, the function $\Xi$ introduced in Eq. \eqref{eq:RTFdecE}, can be decomposed as shown in Eq.~\eqref{eq:delE} where each term in the summation satisfies an integrodifferential equation
    \begin{subequations}\label{eq:intdeltE}
    \begin{align}
      \Xi_{\mathrm{g}}(x_A,\mathbf x_B)&=\frac{R_{AB}}{2}\int_0^1\bigg\{\big(k^{00}-2k^{0i}N^i_{AB}+k^{ij}N^i_{AB}N^j_{AB}\big)_{\widetilde{z}_+(\mu)}\nonumber\\
      &-2\big(k^{0i}-k^{ij}N^j_{AB}\big)_{\widetilde{z}_+(\mu)}\left[\frac{\partial\Xi_{\mathrm{g}}}{\partial x^i}\right]_{(x_A,\mathbf{z}_+(\mu))}+\big(\eta^{ij}+k^{ij}\big)_{\widetilde{z}_+(\mu)}\left[\frac{\partial\Xi_{\mathrm{g}}}{\partial x^i}\frac{\partial\Xi_{\mathrm{g}}}{\partial x^j}\right]_{(x_A,\mathbf{z}_+(\mu))}\bigg\}\dd\mu\text{,}\label{eq:intdeltEg}\\
      \Xi_{\mathrm{r}}(x_A,\mathbf x_B)&=\frac{R_{AB}}{2}\int_0^1\bigg\{\big(\kappa^{00}-2\kappa^{0i}N^i_{AB}+\kappa^{ij}N^i_{AB}N^j_{AB}\big)_{\widetilde{z}_+(\mu)}\nonumber\\
      &-2\big(\kappa^{0i}-\kappa^{ij}N^j_{AB}\big)_{\widetilde{z}_+(\mu)}\left[\frac{\partial\Xi_{\mathrm{r}}}{\partial x^i}\right]_{(x_A,\mathbf{z}_+(\mu))}+\big(\eta^{ij}+\kappa^{ij}\big)_{\widetilde{z}_+(\mu)}\left[\frac{\partial\Xi_{\mathrm{r}}}{\partial x^i}\frac{\partial\Xi_{\mathrm{r}}}{\partial x^j}\right]_{(x_A,\mathbf{z}_+(\mu))}\bigg\}\dd\mu\text{,}\label{eq:intdeltEr}
    \end{align}
    and
    \begin{align}
      \Xi_{\mathrm{gr}}(x_A,\mathbf x_B)&=\frac{R_{AB}}{2}\int_0^1\bigg\{\big(\eta^{ij}+k^{ij}+\kappa^{ij}\big)_{\widetilde{z}_+(\mu)}\left[\frac{\partial\Xi_{\mathrm{gr}}}{\partial x^i}\frac{\partial\Xi_{\mathrm{gr}}}{\partial x^j}\right]_{(x_A,\mathbf{z}_+(\mu))}\nonumber\\
      &-2\big(k^{0i}+\kappa^{0i}-(k^{ij}+\kappa^{ij})N^j_{AB}\big)_{\widetilde{z}_+(\mu)}\left[\frac{\partial\Xi_{\mathrm{gr}}}{\partial x^i}\right]_{(x_A,\mathbf{z}_+(\lambda))}\nonumber\\
      &+\big(k^{ij}\big)_{\widetilde{z}_+(\lambda)}\left[\frac{\partial\Xi_{\mathrm{r}}}{\partial x^i}\frac{\partial\Xi_{\mathrm{r}}}{\partial x^j}\right]_{(x_A,\mathbf{z}_+(\mu))}-2\big(k^{0i}-k^{ij}N^j_{AB}\big)_{\widetilde{z}_+(\mu)}\left[\frac{\partial\Xi_{\mathrm{r}}}{\partial x^i}\right]_{(x_A,\mathbf{z}_+(\mu))}\nonumber\\
      &+\big(\kappa^{ij}\big)_{\widetilde{z}(\lambda)}\left[\frac{\partial\Xi_{\mathrm{g}}}{\partial x^i}\frac{\partial\Xi_{\mathrm{g}}}{\partial x^j}\right]_{(x_A,\mathbf{z}_+(\mu))}-2\big(\kappa^{0i}-\kappa^{ij}N^j_{AB}\big)_{\widetilde{z}_+(\mu)}\left[\frac{\partial\Xi_{\mathrm{g}}}{\partial x^i}\right]_{(x_A,\mathbf{z}_+(\mu))}\nonumber\\
      &+2\big(\eta^{ij}+k^{ij}+\kappa^{ij}\big)_{\widetilde{z}_+(\mu)}\left[\frac{\partial\Xi_{\mathrm{g}}}{\partial x^i}\frac{\partial\Xi_{\mathrm{r}}}{\partial x^j}+\frac{\partial\Xi_{\mathrm{g}}}{\partial x^i}\frac{\partial\Xi_{\mathrm{gr}}}{\partial x^j}+\frac{\partial\Xi_{\mathrm{r}}}{\partial x^i}\frac{\partial\Xi_{\mathrm{gr}}}{\partial x^j}\right]_{(x_A,\mathbf{z}_+(\mu))}\bigg\}\dd\mu\text{.}\label{eq:intdeltEgr}
    \end{align}
    \end{subequations}
    The components of the point event $\widetilde{z}_+(\mu)$ are given by
    \begin{equation}
      \widetilde{z}_+(\mu)=\big(x_A^0+\mu R_{AB}+\Xi(x_A,\mathbf z_+(\mu)),\mathbf z_+(\mu)\big)\text{,}
      \label{eq:tz+m}
    \end{equation}
    where $\mathbf z_+(\mu)$ is defined as in Eq. \eqref{eq:z+i}.
    \label{th:theo3}
  \end{theorem}
\end{widetext}

Theorems \ref{th:theo2} and \ref{th:theo3} generalize Theorems 2 and 3 of \cite{2008CQGra..25n5020T} for the optical spacetime. Indeed, in the limit where refractivity vanishes, that is to say $|\kappa^{\mu\nu}|_{\mathrm{max}}\rightarrow 0$, Theorems 2 and 3 of \cite{2008CQGra..25n5020T} are recovered.

From Eqs.~\eqref{eq:intdeltR}, we see that the choice \eqref{eq:delR} does achieve the separation between the different physical quantities entering the computation of the total time delay. As a matter of fact, the right-hand sides of Eqs.~\eqref{eq:intdeltRg} and \eqref{eq:intdeltRr} contain purely gravitational and purely refractive quantities, respectively. The right-hand side of Eq.~\eqref{eq:intdeltRgr} regroups all terms being a mixture of both.

However, as may be observed from the presence of the total delay in Eq. \eqref{eq:tz-m}, the expressions of the different contributions are not fully independent but remain linked via the path of integration. In the next section, we shall further discuss this point and shall present a recursive resolution method for determining the time delay functions at any order.

\section{General expansions of the time delay functions}
\label{sec:exp}

Because the line integrals in Eqs.~\eqref{eq:intdeltR} are taken along the path $\widetilde{z}_-(\lambda)$ for $0\leqslant\lambda\leqslant1$, the time delay functions $\Delta_{\mathrm{g}}/c$, $\Delta_{\mathrm{r}}/c$, and $\Delta_{\mathrm{gr}}/c$ cannot be solved independently from each other. Indeed, the total delay appearing in Eq.~\eqref{eq:tz-m} depends on the three functions as can be seen from the decomposition \eqref{eq:delR}. Therefore, a systematic and recursive resolution of $\Delta/c$ can only be achieved once the relative contributions of $\Delta_{\mathrm{g}}/c$, $\Delta_{\mathrm{r}}/c$, and $\Delta_{\mathrm{gr}}/c$ to the total time delay are known.

In Sec. \ref{subsec:QM}, we first show how to determine the relative importance between the different contributions. Then within the approximation of a quasi-Minkowskian path, we show that the interdependence between each function $\Delta_{\mathrm{g}}$, $\Delta_{\mathrm{r}}$, and $\Delta_{\mathrm{gr}}$ can always be rejected to the following order during the resolution of $\Delta$. This fact allows one to sort out the occurrence of the different contributions within the determination of the total delay function (cf. Theorems~\ref{th:theo4} and \ref{th:theo5}). In Sec.~\ref{sec:detref}, we assume that the refractive components of the optical metric admit a series expansion in terms of a parameter $N_0$. Then we show that the refractive delay functions can be determined to any order through a recursive resolution method presented in Theorems~\ref{th:theo6} and \ref{th:theo7}. In Sec.~\ref{sec:detgra}, we assume that the gravitational components of the spacetime metric admit a post-Minkowskian expansion (series expansion in ascending power of $G$). Then the recursive method allowing one to determine the gravitational delay expressions up to any order is presented in Theorems~\ref{th:theo8} and \ref{th:theo9}. Finally in Sec.~\ref{sec:detcou}, we determine the coupling delay expressions up to any order within Theorems \ref{th:theo10} and \ref{th:theo11}.

\subsection{Quasi-Minkowskian path regime}
\label{subsec:QM}

As shown in Theorems \ref{th:theo2} and \ref{th:theo3}, the relative magnitude between each contribution to the total time delay rely on the line integrals of the gravitational and the refractive perturbations. Generally speaking, if gravity acts all along the light path $\Gamma$ joining $x_A$ to $x_B$, the refractive domain $\mathcal{D}$ is localized in spacetime and it follows that the action of refractivity remains bounded to a certain portion of $\Gamma$. Therefore, in order to determine the relative contributions of each time delay function, not only the relative magnitude between the gravitational and refractive perturbations must be known, but also the typical length scales over which each perturbation acts. Henceforth, let $\ell$ be the length of $\Gamma$ passing through $\mathcal{D}$. For a Minkowskian path, we always have $\ell\leqslant R_{AB}$ whatever the size of $\mathcal{D}$ is.

From Eq.~\eqref{eq:RTFdecR} which has been formulated under the assumption that the light path is quasi-Minkowskian, we deduce $\Delta/R_{AB}\ll 1$. This implies that $\Delta_{\mathrm{g}}/R_{AB}\ll 1$, $\Delta_{\mathrm{r}}/R_{AB}\ll 1$, and $\Delta_{\mathrm{gr}}/R_{AB}\ll 1$. Considering that $\Delta_{\mathrm{gr}}$ represents the coupling contributions, its magnitude is expected to be of the order
\begin{equation}
  \frac{\Delta_{\mathrm{gr}}}{R_{AB}}\sim\left(\frac{\Delta_{\mathrm{g}}}{R_{AB}}\right)\left(\frac{\Delta_{\mathrm{r}}}{R_{AB}}\right)\text{.}
\end{equation}
Therefore, we can first focus on the relative importance between the gravitational and the refractive contributions.

To do so, let us introduce the parameter $s$ defined by
\begin{equation}
  s=\left\lfloor\frac{\log_{10}(\Delta_{\mathrm{g}}/R_{AB})}{\log_{10}(\Delta_{\mathrm{r}}/R_{AB})}\right\rceil\text{,}
  \label{eq:s}
\end{equation}
with $\lfloor i\rceil$ denoting the operation of rounding to the nearest integer of $i$. Hereafter, we intend to show that the expansion pattern of the delay functions can totally be determined once $s$ is known. Indeed, $s$ allows one to sort out the occurrences of the gravitational and refractive terms in the determination of the total delay functions. 

Because we are only focusing on the main integer value of $s$ in Eq. \eqref{eq:s}, it is sufficient to get the first-order expressions of $\Delta_{\mathrm{g}}$ and $\Delta_{\mathrm{r}}$. Therefore, in Eqs. \eqref{eq:intdeltR}, line integrals can be changed into line integrals along the Minkowskian path between $x_A$ and $x_B$ by performing a Taylor series expansion of $\kappa^{\mu\nu}(\widetilde{z}_-(\lambda))$ and $k^{\mu\nu}(\widetilde{z}_-(\lambda))$ about the point event $z_-(\lambda)$ whose components are given by
\begin{equation}
  z_-(\lambda)=\big(x_B^0-\lambda R_{AB},\mathbf{z}_-(\lambda)\big)\text{.}
  \label{eq:z-m}
\end{equation}
Thus, optical metric components become an infinite series in ascending power of the total time delay
\begin{subequations}\label{eq:Tay}
\begin{align}
  &k^{\mu\nu}\left(\widetilde{z}_-(\lambda),\frac{\Delta(\mathbf{z}_-(\lambda),x_B)}{R_{AB}}\right)=k^{\mu\nu}(z_-(\lambda))\nonumber\\
  &+\sum_{l=1}^{\infty}\frac{(-R_{AB})^l}{l!}\left(\frac{\Delta(\mathbf{z}_-(\lambda),x_B)}{R_{AB}}\right)^l\big[\partial_0^lk^{\mu\nu}\big]_{z_-(\lambda)}\text{,}
  \label{eq:kTay}
\end{align}
and
\begin{align}
  &\kappa^{\mu\nu}\left(\widetilde{z}_-(\lambda),\frac{\Delta(\mathbf{z}_-(\lambda),x_B)}{R_{AB}}\right)=\kappa^{\mu\nu}(z_-(\lambda))\nonumber\\
  &+\sum_{l=1}^{\infty}\frac{(-R_{AB})^l}{l!}\left(\frac{\Delta(\mathbf{z}_-(\lambda),x_B)}{R_{AB}}\right)^{l}\big[\partial_0^l\kappa^{\mu\nu}\big]_{z_-(\lambda)}\text{.}
  \label{eq:kapTay}
\end{align}
\end{subequations}

After inserting these expressions into Eqs. \eqref{eq:intdeltRg} and \eqref{eq:intdeltRr}, we infer that the zeroth-order terms in Eqs.~\eqref{eq:Tay}, namely, $k^{\mu\nu}(z_-(\lambda))$ and $\kappa^{\mu\nu}(z_-(\lambda))$, correspond to the first-order determination of the gravitational and the refractive delays
\begin{subequations}\label{eq:dgdrrelmag}
\begin{equation}
  \frac{\Delta_{\mathrm{g}}^{(1)}}{R_{AB}}=\frac{1}{2}\int_0^1(k^{00}-2k^{0i}N^i_{AB}+k^{ij}N^i_{AB}N^j_{AB})_{z_-(\lambda)}\dd\lambda\text{,}
\end{equation}
and
\begin{equation}
\frac{\Delta_{\mathrm{r}}^{(1)}}{R_{AB}}=\frac{1}{2}\int_0^1(\kappa^{00}-2\kappa^{0i}N^i_{AB}+\kappa^{ij}N^i_{AB}N^j_{AB})_{z_-(\lambda)}\dd\lambda\text{.}
\end{equation}
\end{subequations}
[We will see with Eqs. \eqref{eq:intdelRgPM1} and \eqref{eq:intdelRrPM1} that in the context of a quasi-Minkowskian path, these equations can be further simplified. But for now, let us pursue the discussion with Eqs. \eqref{eq:dgdrrelmag}]. These equations can be inserted into Eq. \eqref{eq:s} in order to determine the value of $s$.

Now, we shall discuss how the expansion pattern of the delay functions can be inferred from $s$. Henceforth, we consider the case $s\in\mathbb{N}_{>1}$ (the result will still be valid for $s\in\mathbb{N}_{>0}$). In other words, we suppose that the refractive perturbation is dominant with respect to the gravitational one \footnote{Results assuming that the gravitational perturbation is the leading term can easily be derived by applying the exact same following approach. However, in that case, it must be noted that $s$ should be introduced switching the numerator and the denominator in the right-hand side of Eq. \eqref{eq:s}.}.

In order to simplify the next discussion, and without loss of generality, we focus on orders of magnitude only. In addition, we consider that the light path occurs in a sufficiently small region of spacetime where the metric components do not vary significantly. Thus, we deduce from Eqs. \eqref{eq:dgdrrelmag} that
\begin{equation}
  \frac{\Delta_{\mathrm{g}}^{(1)}}{R_{AB}}\sim |k^{\mu\nu}|_{\mathrm{max}}\text{,} \qquad \frac{\Delta_{\mathrm{r}}^{(1)}}{R_{AB}}\sim\frac{\ell}{R_{AB}}|\kappa^{\mu\nu}|_{\mathrm{max}}\text{.}
  \label{eq:del0ord}
\end{equation}

In order to keep track of the relative magnitude between the gravitational and the refractive terms, we introduce a dimensionless parameter denoted by $\varepsilon$ and being of the order of the dominant term, that is
\begin{equation}
  \varepsilon=\frac{\Delta_{\mathrm{r}}^{(1)}}{R_{AB}}\text{.}
  \label{eq:varepDr}
\end{equation}
Thus, from Eqs. \eqref{eq:del0ord} and \eqref{eq:s}, we immediately infer
\begin{equation}
  \frac{\Delta_{\mathrm{g}}^{(1)}}{R_{AB}}\sim\mathcal{O}(\varepsilon^s)\text{.}
  \label{eq:varepDg}
\end{equation}
Therefore, the first-order expression of the total delay is driven by the refractive term only
\begin{equation}
  \Delta^{(1)}(\mathbf{x}_A,x_B)=\Delta_{\mathrm{r}}^{(1)}(\mathbf{x}_A,x_B)
  \label{eq:del1delr}
\end{equation}
which means that
\begin{equation}
  \frac{\Delta^{(1)}}{R_{AB}}=\varepsilon\text{,}
  \label{eq:deltot1}
\end{equation}
when $s>1$ in Eq. \eqref{eq:s}.

Let us take a look at the relation between metric components. Equations \eqref{eq:varepDg}, \eqref{eq:varepDr}, and \eqref{eq:del0ord}, allow us to deduce
\begin{equation}
  |k^{\mu\nu}|_{\mathrm{max}}\sim\mathcal{O}(\varepsilon^s)\text{,} \qquad \frac{\ell}{R_{AB}}|\kappa^{\mu\nu}|_{\mathrm{max}}\sim\mathcal{O}(\varepsilon)\text{.}
  \label{eq:kkap0}
\end{equation}
At the same time, it might be seen from Eqs. \eqref{eq:intdeltRr}, \eqref{eq:kTay}, and \eqref{eq:del1delr} that the second-order refractive delay is driven by terms such as
\begin{equation*}
  \frac{\ell}{R_{AB}}|\kappa^{\mu\nu}|_{\mathrm{max}}\frac{\Delta^{(1)}}{R_{AB}}\text{,} \qquad \frac{\Delta^{(1)}}{R_{AB}}\frac{\Delta^{(1)}}{R_{AB}}\text{,}
\end{equation*}
which, according to Eqs. \eqref{eq:deltot1} and \eqref{eq:kkap0}, are of the order of $\varepsilon^2$. Therefore, we conclude that the series expansion of the total delay goes on like
\begin{equation}
  \Delta^{(l)}(\mathbf{x}_A,x_B)=\Delta_{\mathrm{r}}^{(l)}(\mathbf{x}_A,x_B)
\end{equation}
for $1\leqslant l<s$.

The first occurrence of the gravitational contribution to the total delay arises for $l=s$ as anticipated in Eq.~\eqref{eq:varepDg}. Therefore, the $s$th-order expression of the total delay is given by
\begin{equation}
  \Delta^{(s)}(\mathbf{x}_A,x_B)=\Delta_{\mathrm{r}}^{(s)}(\mathbf{x}_A,x_B)+\Delta_{\mathrm{g}}^{(1)}(\mathbf{x}_A,x_B)\text{.}
\end{equation}
Then by looking at the first-order term in Eq. \eqref{eq:kTay}, one might see that the second-order expression of the gravitational delay is proportional to
\begin{equation}
  \Delta_{\mathrm g}^{(2)}\sim|k^{\mu\nu}|_{\mathrm{max}}\frac{\Delta^{(1)}}{R_{AB}}
  \label{eq:delg2}
\end{equation}
which, according to Eqs. \eqref{eq:deltot1} and \eqref{eq:kkap0}, is of the order of $\varepsilon^{s+1}$. Additionally, after inserting Eqs. \eqref{eq:Tay} into \eqref{eq:intdeltRgr}, we infer that the first-order expression of the coupling delay is driven by terms such like
\begin{equation*}
  \frac{\ell}{R_{AB}}|\kappa^{\mu\nu}|_{\mathrm{max}}\frac{\Delta_{\mathrm g}^{(1)}}{R_{AB}}\text{,} \qquad |k^{\mu\nu}|_{\mathrm{max}}\frac{\Delta_{\mathrm{r}}^{(1)}}{R_{AB}}\text{,} \qquad \frac{\Delta_{\mathrm{r}}^{(1)}}{R_{AB}}\frac{\Delta_{\mathrm g}^{(1)}}{R_{AB}}\text{,}
\end{equation*}
which are of the order of $\varepsilon^{s+1}$ too. Therefore, the $(s+1)$th-order expression of the total delay is given by
\begin{align}
  \Delta^{(s+1)}(\mathbf{x}_A,x_B)&=\Delta_{\mathrm{r}}^{(s+1)}(\mathbf{x}_A,x_B)+\Delta_{\mathrm{g}}^{(2)}(\mathbf{x}_A,x_B)\nonumber\\
   &+\Delta_{\mathrm{gr}}^{(1)}(\mathbf{x}_A,x_B)\text{.}
\end{align}

A quick look at the second-order expression of the coupling delay shows that it is driven by terms proportional to $\varepsilon^{s+2}$. Consequently, one deduces that
\begin{align}
  \Delta^{(l)}(\mathbf{x}_A,x_B)&=\Delta_{\mathrm{r}}^{(l)}(\mathbf{x}_A,x_B)+\Delta_{\mathrm{g}}^{(l-s+1)}(\mathbf{x}_A,x_B)\nonumber\\
   &+\Delta_{\mathrm{gr}}^{(l-s)}(\mathbf{x}_A,x_B)\text{,}
\end{align}
for $l\geqslant s+1$.

To sum up, within the quasi-Minkowskian regime, the total delay satisfies $\Delta/R_{AB}\ll 1$, so the line integrals in Eqs. \eqref{eq:intdeltR} are simplified into line integrals along the Minkowskian path by performing a Taylor series expansion about the point event $z_-(\lambda)$. Then by considering the case where the refractivity is the dominant effect all along the light path $\Gamma$, it results that, in general, the total time delay admits an expansion as follows:
\begin{equation}
  \Delta(\mathbf{x}_A,x_B)=\sum_{l=1}^{\infty}\Delta^{(l)}(\mathbf{x}_A,x_B)
  \label{eq:delRPM}
\end{equation}
where the terms $\Delta^{(l)}$ are proportional to $\varepsilon^lR_{AB}$.

In that respect, the different contributions to the total delay, namely, the refractive, the gravitational, and the coupling delays, all admit series expansion as follows:
\begin{subequations}\label{eq:delRrgrPM}
\begin{align}
  \Delta_{\mathrm{r}}(\mathbf{x}_A,x_B)&=\sum_{l=1}^{\infty}\Delta_{\mathrm{r}}^{(l)}(\mathbf{x}_A,x_B)\text{,}\label{eq:delRrPM}\\
  \Delta_{\mathrm{g}}(\mathbf{x}_A,x_B)&=\sum_{l=1}^{\infty}\Delta_{\mathrm{g}}^{(l)}(\mathbf{x}_A,x_B)\text{,}\label{eq:delRgPM}
\end{align}
and
\begin{equation}
  \Delta_{\mathrm{gr}}(\mathbf{x}_A,x_B)=\sum_{l=1}^{\infty}\Delta_{\mathrm{gr}}^{(l)}(\mathbf{x}_A,x_B)
  \label{eq:delRgrPM}
\end{equation}
\end{subequations}
where the terms $\Delta_{\mathrm{r}}^{(l)}$, $\Delta_{\mathrm{g}}^{(l)}$, and $\Delta_{\mathrm{gr}}^{(l)}$ are of the order of
\begin{equation}
  \frac{\Delta_{\mathrm{r}}^{(l)}}{R_{AB}}\sim\mathcal{O}(\varepsilon^{l})\text{,} \quad \frac{\Delta_{\mathrm{g}}^{(l)}}{R_{AB}}\sim\mathcal{O}(\varepsilon^{l+s-1})\text{,} \quad \frac{\Delta_{\mathrm{gr}}^{(l)}}{R_{AB}}\sim\mathcal{O}(\varepsilon^{l+s})\text{.}
\end{equation}

We recall that $\varepsilon$ is of the order of $\ell/R_{AB}|\kappa^{\mu\nu}|_{\mathrm{max}}$ only for a light path occurring in a sufficiently small region of spacetime where the metric components do not vary significantly. In general, it is given by Eq. \eqref{eq:varepDr}.

By making use of the Heaviside step function
\begin{equation}
  \Theta(i)=\left\{
  \begin{array}{l l}
    1 & \text{for }i\geqslant 0\text{,}\\
    0 & \text{otherwise,}
  \end{array}
  \right.
  \label{eq:stp}
\end{equation}
we can write the terms $\Delta^{(l)}$ in Eq. \eqref{eq:delRPM} as
\begin{align}
  \Delta^{(l)}(\mathbf{x}_A,x_B)&=\Delta_{\mathrm{r}}^{(l)}(\mathbf{x}_A,x_B)+\Theta(l\!-\!s)\Delta_{\mathrm{g}}^{(l-s+1)}(\mathbf{x}_A,x_B)\nonumber\\
  &+\Theta(l\!-\!s\!-\!1)\Delta_{\mathrm{gr}}^{(l-s)}(\mathbf{x}_A,x_B)\text{.}\label{eq:delRPMl}
\end{align}

In the next two sections, according to the fact that the light ray follows a quasi-Minkowskian path, we will assume that the components $\kappa^{\mu\nu}$ and $k^{\mu\nu}$ admit series expansion in ascending power of parameters $N_0$ and $G$, respectively [see Eqs.~\eqref{eq:kapPM} and \eqref{eq:kPM}]. If this fact does not change the pattern of the series expansions \eqref{eq:delRPM} and \eqref{eq:delRrgrPM}, we should nevertheless, for completeness, specify that the quasi-Minkowskian path is parametrized by the expansion coefficients $N_0$ and $G$. Therefore, by making use of Eq. \eqref{eq:delRPMl}, we can state a theorem as follows.

\begin{widetext}
  \begin{theorem}
    Within the quasi-Minkowskian path approximation, when the light path is parametrized by $N_0$ and $G$, the function $\Delta$ admits a series expansion as follows:
    \begin{equation}
      \Delta(\mathbf{x}_A,x_B,N_0,G)=\sum_{l=1}^{\infty}\Delta^{(l)}(\mathbf{x}_A,x_B)\text{,}
      \label{eq:delRPMN0G}
    \end{equation}
    with
    \begin{equation}
      \Delta^{(l)}(\mathbf{x}_A,x_B)=\Delta_{\mathrm{r}}^{(l)}(\mathbf{x}_A,x_B)+\Theta(l-s)\Delta_{\mathrm{g}}^{(l-s+1)}(\mathbf{x}_A,x_B)+\Theta(l-s-1)\Delta_{\mathrm{gr}}^{(l-s)}(\mathbf{x}_A,x_B)\text{.}
    \end{equation}
    The parameter $s\in\mathbb{N}_{>0}$ is determined from Eq. \eqref{eq:s} by making use of the first-order expressions \eqref{eq:intdelRrPM1} and \eqref{eq:intdelRgPM1}.
    \label{th:theo4}
  \end{theorem}
\end{widetext}

A similar reasoning works for the emission time delay function as well. Indeed, the line integrals in Eqs. \eqref{eq:intdeltE} can be Taylor expanded about the point event $z_+(\mu)$ whose components are given by
\begin{equation}
  z_+(\mu)=\big(x_A^0+\mu R_{AB},\mathbf{z}_+(\mu)\big)\text{.}
  \label{eq:z+m}
\end{equation}
Then $\kappa^{\mu\nu}(\widetilde{z}_+(\mu))$ and $k^{\mu\nu}(\widetilde{z}_+(\mu))$ become an infinite series in ascending power of $\Xi$ similarly to what has been done in Eqs. \eqref{eq:Tay}.

Therefore, we end up with a similar expansion for $\Xi$ than for $\Delta$ and we state the following theorem.

\begin{widetext}
  \begin{theorem}
    Within the quasi-Minkowskian path approximation, when the light path is parametrized by $N_0$ and $G$, the function $\Xi$ admits a series expansion as follows:
    \begin{equation}
      \Xi(x_A,\mathbf{x}_B,N_0,G)=\sum_{l=1}^{\infty}\Xi^{(l)}(x_A,\mathbf{x}_B)\text{,}
      \label{eq:delEPMN0G}
    \end{equation}
    with
    \begin{equation}
      \Xi^{(l)}(x_A,\mathbf{x}_B)=\Xi_{\mathrm{r}}^{(l)}(x_A,\mathbf{x}_B)+\Theta(l-s)\Xi_{\mathrm{g}}^{(l-s+1)}(x_A,\mathbf{x}_B)+\Theta(l-s-1)\Xi_{\mathrm{gr}}^{(l-s)}(x_A,\mathbf{x}_B)\text{.}
    \end{equation}
    The parameter $s\in\mathbb{N}_{>0}$ is determined from Eq. \eqref{eq:s} by making use of the first-order expressions \eqref{eq:intdelRrPM1} and \eqref{eq:intdelRgPM1}.
    \label{th:theo5}
  \end{theorem}
\end{widetext}

Equipped with Theorems \ref{th:theo2} to \ref{th:theo5}, we can now recursively determine the integral form of each time delay function in Eqs. \eqref{eq:del}.

\subsection{The refractive time delay functions}
\label{sec:detref}

We saw in Sec.~\ref{sec:intTF} [cf. Eqs. \eqref{eq:approxPM}], that a quasi-Minkowskian path implies small refractivities, that is to say $N(x)\ll1$. Let $N_0=N(x_0)$ be the refractivity at a well chosen point event $x_0\in\mathcal{D}$ located on $\Gamma$.

If $x_0$ is chosen so that $N_0$ is the maximum value of the refractivity along $\Gamma$, we can always write
\begin{equation}
  N(x)=N_0\left(\frac{N(x)}{N_0}\right)\text{,} \qquad N(x)\leqslant N_0\text{.}
\end{equation}
Hence, considering a quasi-Minkowskian light path, it follows that $N_0\ll1$. Therefore, we can always expand the components $\kappa^{\mu\nu}$ in ascending power of $N_0$ such as
\begin{equation}
  \kappa^{\mu\nu}(x,N_0)=\sum_{l=1}^{\infty}\kappa^{\mu\nu}_{(l)}(x)
  \label{eq:kapPM}
\end{equation}
where $\kappa^{\mu\nu}_{(l)}\propto(N_0)^l$.

Considering that the wave vector is by definition a covector [see Eq. \eqref{eq:kdef}], the optical metric is intrinsically defined for its contravariant components as seen from Eq.~\eqref{eq:beik}. Therefore, the covariant components of the optical metric are not needed \emph{a priori} to solve the time and frequency transfers. However, we provide their expressions in Sec. \ref{sec:genexp} for completeness.

As discussed previously, line integrations in Eq.~\eqref{eq:intdeltRr} are taken along the real light path $\widetilde{z}_-(\lambda)$ for $0\leqslant\lambda\leqslant 1$. Within the quasi-Minkowskian path regime, we saw in Eq. \eqref{eq:kapTay} that the metric components $\kappa^{\mu\nu}(\widetilde{z}_-(\lambda))$ can be expanded in ascending power of the total delay. In that respect, the right-hand side of Eq.~\eqref{eq:kapTay} involves terms such as $\kappa^{\mu\nu}(z_-(\lambda))$, where $z_-(\lambda)$ is given in Eq. \eqref{eq:z-m}. By making use of Eq. \eqref{eq:kapPM}, we immediately find
\begin{equation}
  \kappa^{\mu\nu}(z_-(\lambda),N_0)=\sum_{l=1}^{\infty}\kappa^{\mu\nu}_{(l)}(z_-(\lambda))\text{.}
  \label{eq:kapPMz-}
\end{equation}

Therefore, the general expansion of $\kappa^{\mu\nu}(\widetilde{z}_-(\lambda))$ is obtained after substituting for $\Delta$ and $\kappa^{\mu\nu}$ from Eqs.~\eqref{eq:delRPMN0G} and \eqref{eq:kapPMz-} into \eqref{eq:kapTay}, respectively. After some algebra, we find a relation as follows
\begin{equation}
  \kappa^{\mu\nu}(\widetilde{z}_-(\lambda),N_0,G)=\sum_{l=1}^{\infty}\widehat{\kappa}^{\mu\nu}_{-(l)}(z_-(\lambda),x_B)
  \label{eq:kapRPM}
\end{equation}
where the quantities $\widehat{\kappa}^{\mu\nu}_{-(l)}(z_-(\lambda),x_B)$ are given by
\begin{subequations}\label{eq:kapRPMdec}
\begin{equation}
  \widehat{\kappa}^{\mu\nu}_{-(1)}(z_-(\lambda),x_B)=\kappa^{\mu\nu}_{(1)}(z_-(\lambda))\text{,}
\end{equation}
and
\begin{align}
  &\widehat{\kappa}^{\mu\nu}_{-(l)}(z_-(\lambda),x_B)=\kappa^{\mu\nu}_{(l)}(z_-(\lambda))\nonumber\\
  &+\sum_{m=1}^{l-1}\sum_{n=1}^{m}\Phi_-^{(m,n)}(\mathbf z_-(\lambda),x_B)\left[\frac{\partial^n\kappa^{\mu\nu}_{(l-m)}}{(\partial x^0)^n}\right]_{z_-(\lambda)}
\end{align}
\end{subequations}
for $l\geqslant 2$. The function $\Phi_-^{(l,m)}(\mathbf x,x_B)$, with $l\geqslant 1$ and $1\leqslant m\leqslant l$, is called a reception function \cite{2008CQGra..25n5020T} and is defined such that
\begin{align}
  \Phi_-^{(l,m)}&(\mathbf x,x_B)=\frac{(-1)^m}{m!}\times\nonumber\\
  &\sum_{n_1+\cdots+n_m=l-m}\Bigg[\prod_{d=1}^m\Delta^{(n_d+1)}(\mathbf x,x_B)\Bigg]\text{,}\label{eq:recfun}
\end{align}
with $n_1,\ldots,n_m\in\mathbb{N}_{\geqslant 0}$. The summation in \eqref{eq:recfun} is taken over all sequences of $n_1$ through $n_m$ such that the sum of all $n_m$ is $l-m$.

Finally, by substituting for $\kappa^{\mu\nu}(\widetilde{z}_-(\lambda))$ from Eq. \eqref{eq:kapRPM} into \eqref{eq:intdeltRr}, we infer the theorem which follows concerning the refractive time delay function at reception.

\begin{widetext}
  \begin{theorem}
    In the optical spacetime, within the quasi-Minkowskian path approximation, $\Delta$ admits the series expansion introduced in Theorem \ref{th:theo4}, so the function $\Delta_{\mathrm{r}}$ is given by
    \begin{equation}
      \Delta_{\mathrm{r}}(\mathbf{x}_A,x_B,N_0,G)=\sum_{l=1}^{\infty}\Delta_{\mathrm{r}}^{(l)}(\mathbf{x}_A,x_B)
      \label{eq:delRrPMN0G}
    \end{equation}
    where
    \begin{subequations}\label{eq:intdelRrPM}
    \begin{align}
      \Delta_{\mathrm{r}}^{(1)}(\mathbf{x}_A,x_B)&=\frac{R_{AB}}{2}\int^{1}_{0}\big(\kappa^{00}_{(1)}-2\kappa^{0i}_{(1)}N^i_{AB}+\kappa^{ij}_{(1)}N^i_{AB}N^j_{AB}\big)_{z_-(\lambda)}\dd\lambda\text{,}\label{eq:intdelRrPM1}\\
      \Delta_{\mathrm{r}}^{(2)}(\mathbf{x}_A,x_B)&=\frac{R_{AB}}{2}\int^{1}_{0}\Bigg\{\big(\widehat{\kappa}^{00}_{-(2)}-2\widehat{\kappa}^{0i}_{-(2)}N^i_{AB}+\widehat{\kappa}^{ij}_{-(2)}N^i_{AB}N^j_{AB}\big)_{(z_-(\lambda),x_B)}\nonumber\\
      &+2\big(\kappa^{0i}_{(1)}-\kappa^{ij}_{(1)}N_{AB}^j\big)_{z_-(\lambda)}\left[\frac{\partial\Delta_{\mathrm{r}}^{(1)}}{\partial x^i}\right]_{(\mathbf{z}_-(\lambda),x_B)}+\eta^{ij}\left[\frac{\partial\Delta_{\mathrm{r}}^{(1)}}{\partial x^i}\frac{\partial\Delta_{\mathrm{r}}^{(1)}}{\partial x^j}\right]_{(\mathbf{z}_-(\lambda),x_B)}\Bigg\}\dd\lambda\text{,}
    \end{align}
    and
    \begin{align}
      \Delta_{\mathrm{r}}^{(l)}(\mathbf{x}_A,x_B)&=\frac{R_{AB}}{2}\int^{1}_{0}\Bigg\{\big(\widehat{\kappa}^{00}_{-(l)}-2\widehat{\kappa}^{0i}_{-(l)}N^i_{AB}+\widehat{\kappa}^{ij}_{-(l)}N^i_{AB}N^j_{AB}\big)_{(z_-(\lambda),x_B)}\nonumber\\
      &+2\sum_{m=1}^{l-1}\big(\widehat\kappa^{0i}_{-(m)}-\widehat\kappa^{ij}_{-(m)}N_{AB}^j\big)_{(z_-(\lambda),x_B)}\left[\frac{\partial\Delta_{\mathrm{r}}^{(l-m)}}{\partial x^i}\right]_{(\mathbf{z}_-(\lambda),x_B)}+\eta^{ij}\sum_{m=1}^{l-1}\left[\frac{\partial\Delta_{\mathrm{r}}^{(m)}}{\partial x^i}\frac{\partial\Delta_{\mathrm{r}}^{(l-m)}}{\partial x^j}\right]_{(\mathbf{z}_-(\lambda),x_B)}\nonumber\\
      &+\sum_{m=1}^{l-2}\big(\widehat{\kappa}^{ij}_{-(m)}\big)_{(z_-(\lambda),x_B)}\sum_{n=1}^{l-m-1}\left[\frac{\partial\Delta_{\mathrm{r}}^{(n)}}{\partial x^i}\frac{\partial\Delta_{\mathrm{r}}^{(l-m-n)}}{\partial x^j}\right]_{(\mathbf{z}_-(\lambda),x_B)}\Bigg\}\dd\lambda
    \end{align}
    \end{subequations}
    for $l \geqslant 3$. The quantities $\widehat{\kappa}^{\mu\nu}_{-(l)}(z_-(\lambda),x_B)$ are defined in Eqs. \eqref{eq:kapRPMdec}.
    \label{th:theo6}
  \end{theorem}
\end{widetext}

Applying the exact same reasoning, the analogous theorem for the refractive time delay function at emission can be stated as well. Line integrations in Eq.~\eqref{eq:intdeltEr} are taken along the light path $\widetilde{z}_+(\mu)$ for $0\leqslant\mu\leqslant 1$. After Taylor expanding the light path about the point event $z_+(\mu)$ [cf. Eq. \eqref{eq:z+m}], the right-hand side of Eq.~\eqref{eq:kapTay} involves terms such as $\kappa^{\mu\nu}(z_+(\mu))$. After making use of Eq.~\eqref{eq:kapPM}, we find
\begin{equation}
  \kappa^{\mu\nu}(z_+(\mu),N_0)=\sum_{l=1}^{\infty}\kappa^{\mu\nu}_{(l)}(z_+(\mu))\text{.}
  \label{eq:kapPMz+}  
\end{equation}

The general expansion of $\kappa^{\mu\nu}(\widetilde{z}_+(\mu))$ is obtained after substituting for $\Xi$ from Eq.~\eqref{eq:delEPMN0G} and for $\kappa^{\mu\nu}$ from Eq.~\eqref{eq:kapPMz+}, into \eqref{eq:kapTay}. After some algebra, we find
\begin{equation}
  \kappa^{\mu\nu}(\widetilde{z}_+(\mu),N_0,G)=\sum_{l=1}^{\infty}\widehat{\kappa}^{\mu\nu}_{+(l)}(x_A,z_+(\mu))
  \label{eq:kapEPM}
\end{equation}
where the quantities $\widehat{\kappa}^{\mu\nu}_{+(l)}(x_A,z_+(\mu))$ are given by
\begin{subequations}\label{eq:kapEPMdec}
\begin{equation}
  \widehat{\kappa}^{\mu\nu}_{+(1)}(x_A,z_+(\mu))=\kappa^{\mu\nu}_{(1)}(z_+(\mu))\text{,}
\end{equation}
and
\begin{align}
  &\widehat{\kappa}^{\mu\nu}_{+(l)}(x_A,z_+(\mu))=\kappa^{\mu\nu}_{(l)}(z_+(\mu))\nonumber\\
  &+\sum_{m=1}^{l-1}\sum_{n=1}^{m}\Phi_+^{(m,n)}(x_A,\mathbf z_+(\mu))\left[\frac{\partial^n\kappa^{\mu\nu}_{(l-m)}}{(\partial x^0)^n}\right]_{z_+(\mu)}
\end{align}
\end{subequations}
for $l\geqslant 2$. The function $\Phi_+^{(l,m)}(x_A,\mathbf x)$, with $l\geqslant 1$ and $1\leqslant m\leqslant l$, is called an emission function \cite{2008CQGra..25n5020T} and is defined such that
\begin{align}
  \Phi_+^{(l,m)}&(x_A,\mathbf x)=\frac{1}{m!}\times\nonumber\\
  &\sum_{n_1+\cdots+n_m=l-m}\Bigg[\prod_{d=1}^m\Xi^{(n_d+1)}(x_A,\mathbf x)\Bigg]\text{,}\label{eq:emifun}
\end{align}
with $n_1,\ldots,n_m\in\mathbb{N}_{\geqslant 0}$. The summation in Eq. \eqref{eq:emifun} is taken over all sequences of $n_1$ through $n_m$ such that the sum of all $n_m$ is $l-m$.

Finally, the theorem for the refractive time delay function at emission is obtained after substituting for $\kappa^{\mu\nu}(\widetilde{z}_+(\mu))$ from Eq. \eqref{eq:kapEPM} into \eqref{eq:intdeltEr}.

\begin{widetext}
  \begin{theorem}
    In the optical spacetime, within the quasi-Minkowskian path approximation, $\Xi$ admits the series expansion introduced in Theorem \ref{th:theo5}, so the function $\Xi_{\mathrm{r}}$ is given by
    \begin{equation}
      \Xi_{\mathrm{r}}(x_A,\mathbf{x}_B,N_0,G)=\sum_{l=1}^{\infty}\Xi_{\mathrm{r}}^{(l)}(x_A,\mathbf{x}_B)
      \label{eq:delErPM}
    \end{equation}
    where
    \begin{subequations}\label{eq:intdelErPM}
    \begin{align}
      \Xi_{\mathrm{r}}^{(1)}(x_A,\mathbf{x}_B)&=\frac{R_{AB}}{2}\int^{1}_{0}\big(\kappa^{00}_{(1)}-2\kappa^{0i}_{(1)}N^i_{AB}+\kappa^{ij}_{(1)}N^i_{AB}N^j_{AB}\big)_{z_+(\mu)}\dd\mu\text{,}\\
      \Xi_{\mathrm{r}}^{(2)}(x_A,\mathbf{x}_B)&=\frac{R_{AB}}{2}\int^{1}_{0}\Bigg\{\big(\widehat{\kappa}^{00}_{+(2)}-2\widehat{\kappa}^{0i}_{+(2)}N^i_{AB}+\widehat{\kappa}^{ij}_{+(2)}N^i_{AB}N^j_{AB}\big)_{(x_A,z_+(\mu))}\nonumber\\
      &-2\big(\kappa^{0i}_{(1)}-\kappa^{ij}_{(1)}N_{AB}^j\big)_{z_+(\mu)}\left[\frac{\partial\Xi_{\mathrm{r}}^{(1)}}{\partial x^i}\right]_{(x_A,\mathbf{z}_+(\mu))}+\eta^{ij}\left[\frac{\partial\Xi_{\mathrm{r}}^{(1)}}{\partial x^i}\frac{\partial\Xi_{\mathrm{r}}^{(1)}}{\partial x^j}\right]_{(x_A,\mathbf{z}_+(\mu))}\Bigg\}\dd\mu\text{,}
    \end{align}
    and
    \begin{align}
      \Xi_{\mathrm{r}}^{(l)}(x_A,\mathbf{x}_B)&=\frac{R_{AB}}{2}\int^{1}_{0}\Bigg\{\big(\widehat{\kappa}^{00}_{+(l)}-2\widehat{\kappa}^{0i}_{+(l)}N^i_{AB}+\widehat{\kappa}^{ij}_{+(l)}N^i_{AB}N^j_{AB}\big)_{(x_A,z_+(\mu))}\nonumber\\
      &-2\sum_{m=1}^{l-1}\big(\widehat\kappa^{0i}_{+(m)}-\widehat\kappa^{ij}_{+(m)}N_{AB}^j\big)_{(x_A,z_+(\mu))}\left[\frac{\partial\Xi_{\mathrm{r}}^{(l-m)}}{\partial x^i}\right]_{(x_A,\mathbf{z}_+(\mu))}+\eta^{ij}\sum_{m=1}^{l-1}\left[\frac{\partial\Xi_{\mathrm{r}}^{(m)}}{\partial x^i}\frac{\partial\Xi_{\mathrm{r}}^{(l-m)}}{\partial x^j}\right]_{(x_A,\mathbf{z}_+(\mu))}\nonumber\\
      &+\sum_{m=1}^{l-2}\big(\widehat{\kappa}^{ij}_{+(m)}\big)_{(x_A,z_+(\mu))}\sum_{n=1}^{l-m-1}\left[\frac{\partial\Xi_{\mathrm{r}}^{(n)}}{\partial x^i}\frac{\partial\Xi_{\mathrm{r}}^{(l-m-n)}}{\partial x^j}\right]_{(x_A,\mathbf{z}_+(\mu))}\Bigg\}\dd\mu
    \end{align}
    \end{subequations}
    for $l \geqslant 3$. The quantities $\widehat{\kappa}^{\mu\nu}_{+(l)}(x_A,z_+(\mu))$ are defined in Eqs. \eqref{eq:kapEPMdec}.
    \label{th:theo7}
  \end{theorem}
\end{widetext}

In the case where the components $\kappa^{\mu\nu}$ represent the leading perturbation [see Eq. \eqref{eq:s}], it may be seen that the expansion pattern in Theorems \ref{th:theo6} and \ref{th:theo7} is almost the same as the one in Theorems 4 and 5 of \cite{2008CQGra..25n5020T}. Actually, if one assumes that the gravitational components $k^{\mu\nu}$ are the leading perturbations, one obtains quasisimilar theorems than Theorems 4 and 5 of \cite{2008CQGra..25n5020T} (the difference would be in the definition of the quantities $\widehat{k}^{\mu\nu}_{+(l)}$).

\subsection{The gravitational time delay functions}
\label{sec:detgra}

Following \cite{2008CQGra..25n5020T}, we suppose that the gravitational perturbation terms $h_{\mu\nu}$ can be expressed as a post-Minkowskian expansion, such as
\begin{equation}
  h_{\mu\nu}(x,G)=\sum_{l=1}^{\infty}h_{\mu\nu}^{(l)}(x)\text{.}
  \label{eq:hPM}
\end{equation}
The contravariant components are given by
\begin{equation}
  k^{\mu\nu}(x,G)=\sum_{l=1}^{\infty}k^{\mu\nu}_{(l)}(x)
  \label{eq:kPM}
\end{equation}
where the components $k^{\mu\nu}_{(l)}$ can be recursively determined using the following relationships:
\begin{subequations}\label{eq:kPMord}
\begin{align}
  k^{\mu\nu}_{(1)}&=-\eta^{\mu\alpha}\eta^{\beta\nu}h_{\alpha\beta}^{(1)}\text{,}\label{eq:kPMord1}\\
  k^{\mu\nu}_{(l)}&=-\eta^{\mu\alpha}\eta^{\beta\nu}h_{\alpha\beta}^{(l)}-\sum_{m=1}^{l-1}\eta^{\mu\alpha}h_{\alpha\beta}^{(m)}k^{\beta\nu}_{(l-m)}\label{eq:kPMord2}
\end{align}
\end{subequations}
for $l\geqslant 2$.

The right-hand side of Eq.~\eqref{eq:kTay} involves terms such as $k^{\mu\nu}(z_-(\lambda))$. Thus, by making use of Eq. \eqref{eq:kPM}, we find
 \begin{equation}
  k^{\mu\nu}(z_-(\lambda),G)=\sum_{l=1}^{\infty}k^{\mu\nu}_{(l)}(z_-(\lambda))
  \label{eq:kPMz-}
\end{equation}
where $k^{\mu\nu}_{(l)}\propto G^l$.

Therefore, the general expansion of $k^{\mu\nu}(\widetilde{z}_-(\lambda))$ is obtained after substituting for $\Delta$ and $k^{\mu\nu}$ from Eqs.~\eqref{eq:delRPMN0G} and \eqref{eq:kPMz-} into \eqref{eq:kTay}, respectively. After some algebra, we find the following expression:
\begin{equation}
  k^{\mu\nu}(\widetilde{z}_-(\lambda),N_0,G)=\sum_{l=s}^{\infty}\widehat{k}^{\mu\nu}_{-(l)}(z_-(\lambda),x_B)
  \label{eq:kRPM}
\end{equation}
where the quantities $\widehat{k}^{\mu\nu}_{-(l)}(z_-(\lambda),x_B)$ are given for $l\geqslant s$ by the following expressions:
\begin{subequations}\label{eq:kRPMdec}
\begin{equation}
  \widehat{k}^{\mu\nu}_{-(s)}(z_-(\lambda),x_B)=k_{(1)}^{\mu\nu}(z_-(\lambda))\text{,}
\end{equation}
and
\begin{align}
  &\widehat{k}^{\mu\nu}_{-(ps)}(z_-(\lambda),x_B)=k^{\mu\nu}_{(p)}(z_-(\lambda))\nonumber\\
  &+\sum_{m=1}^{p-1}\sum_{n=1}^{ms}\Phi_-^{(ms,n)}(\mathbf z_-(\lambda),x_B)\left[\frac{\partial^nk_{(p-m)}^{\mu\nu}}{(\partial x^0)^n}\right]_{z_-(\lambda)}\label{eq:kRPMdecl}
\end{align}
for $p\geqslant 2$, and
\begin{align}
  &\widehat{k}^{\mu\nu}_{-(ps+q)}(z_-(\lambda),x_B)=\nonumber\\
  &+\sum_{n=1}^{q}\Phi_-^{(q,n)}(\mathbf z_-(\lambda),x_B)\left[\frac{\partial^nk_{(p)}^{\mu\nu}}{(\partial x^0)^n}\right]_{z_-(\lambda)}\nonumber\\
  &+\sum_{m=1}^{p-1}\sum_{n=1}^{ms+q}\Phi_-^{(ms+q,n)}(\mathbf z_-(\lambda),x_B)\left[\frac{\partial^nk_{(p-m)}^{\mu\nu}}{(\partial x^0)^n}\right]_{z_-(\lambda)}\label{eq:kRPMdecl}
\end{align}
\end{subequations}
for $p\geqslant 1$ and $1\leqslant q\leqslant s-1$, where $p$ and $q$ are determined from $l$ using the following relationships:
\begin{align}
  p&=\lfloor l/s\rfloor\text{,} & q&=l-ps\text{,}\label{eq:pq}
\end{align}
with $\lfloor i\rfloor$ denoting the integer part of $i$.

By substituting for $k^{\mu\nu}(\widetilde{z}_-(\lambda))$ from Eq. \eqref{eq:kRPM} into \eqref{eq:intdeltRg}, we infer the theorem which follows concerning the gravitational time delay function at reception.

\begin{widetext}
  \begin{theorem}
    In the optical spacetime, within the quasi-Minkowskian path approximation, $\Delta$ admits the series expansion introduced in Theorem \ref{th:theo4}, so the function $\Delta_{\mathrm{g}}$ is given by
    \begin{equation}
      \Delta_{\mathrm{g}}(\mathbf{x}_A,x_B,N_0,G)=\sum_{l=1}^{\infty}\Delta_{\mathrm{g}}^{(l)}(\mathbf{x}_A,x_B)
      \label{eq:delRgPMN0G}
    \end{equation}
    where
    \begin{subequations}\label{eq:intdelRgPM}
    \begin{align}
      \Delta_{\mathrm{g}}^{(1)}(\mathbf{x}_A,x_B)&=\frac{R_{AB}}{2}\int^{1}_{0}\big(k^{00}_{(1)}-2k^{0i}_{(1)}N^i_{AB}+k^{ij}_{(1)}N^i_{AB}N^j_{AB}\big)_{z_-(\lambda)}\dd\lambda\text{,}\label{eq:intdelRgPM1}\\
      \Delta_{\mathrm{g}}^{(l)}(\mathbf{x}_A,x_B)&=\frac{R_{AB}}{2}\int^{1}_{0}\big(\widehat k^{00}_{-(s+l-1)}-2\widehat k^{0i}_{-(s+l-1)}N^i_{AB}+\widehat k^{ij}_{-(s+l-1)}N^i_{AB}N^j_{AB}\big)_{(z_-(\lambda),x_B)}\dd\lambda
    \end{align}
    for $2\leqslant l\leqslant s$, and
    \begin{align}
      \Delta_{\mathrm{g}}^{(l)}(\mathbf{x}_A,x_B)&=\frac{R_{AB}}{2}\int^{1}_{0}\Bigg\{\big(\widehat{k}^{00}_{-(s+l-1)}-2\widehat{k}^{0i}_{-(s+l-1)}N^i_{AB}+\widehat{k}^{ij}_{-(s+l-1)}N^i_{AB}N^j_{AB}\big)_{(z_-(\lambda),x_B)}\nonumber\\
      &+2\sum_{m=s}^{l-1}\big(\widehat k^{0i}_{-(m)}-\widehat k^{ij}_{-(m)}N_{AB}^j\big)_{(z_-(\lambda),x_B)}\left[\frac{\partial\Delta_{\mathrm{g}}^{(l-m)}}{\partial x^i}\right]_{(\mathbf{z}_-(\lambda),x_B)}\nonumber\\
      &+\eta^{ij}\sum_{m=1}^{l-s}\left[\frac{\partial\Delta_{\mathrm{g}}^{(m)}}{\partial x^i}\frac{\partial\Delta_{\mathrm{g}}^{(l-s-m+1)}}{\partial x^j}\right]_{(\mathbf{z}_-(\lambda),x_B)}\Bigg\}\dd\lambda
    \end{align}
    for $s+1\leqslant l\leqslant 2s$, and finally
    \begin{align}
      \Delta_{\mathrm{g}}^{(l)}(\mathbf{x}_A,x_B)&=\frac{R_{AB}}{2}\int^{1}_{0}\Bigg\{\big(\widehat{k}^{00}_{-(s+l-1)}-2\widehat{k}^{0i}_{-(s+l-1)}N^i_{AB}+\widehat{k}^{ij}_{-(s+l-1)}N^i_{AB}N^j_{AB}\big)_{(z_-(\lambda),x_B)}\nonumber\\
      &+2\sum_{m=s}^{l-1}\big(\widehat k^{0i}_{-(m)}-\widehat k^{ij}_{-(m)}N_{AB}^j\big)_{(z_-(\lambda),x_B)}\left[\frac{\partial\Delta_{\mathrm{g}}^{(l-m)}}{\partial x^i}\right]_{(\mathbf{z}_-(\lambda),x_B)}\nonumber\\
      &+\eta^{ij}\sum_{m=1}^{l-s}\left[\frac{\partial\Delta_{\mathrm{g}}^{(m)}}{\partial x^i}\frac{\partial\Delta_{\mathrm{g}}^{(l-s-m+1)}}{\partial x^j}\right]_{(\mathbf{z}_-(\lambda),x_B)}\nonumber\\
      &+\sum_{m=s}^{l-s-1}\big(\widehat{k}^{ij}_{-(m)}\big)_{(z_-(\lambda),x_B)}\sum_{n=1}^{l-s-m}\left[\frac{\partial\Delta_{\mathrm{g}}^{(n)}}{\partial x^i}\frac{\partial\Delta_{\mathrm{g}}^{(l-s-m-n+1)}}{\partial x^j}\right]_{(\mathbf{z}_-(\lambda),x_B)}\Bigg\}\dd\lambda
    \end{align}
    \end{subequations}
    for $l\geqslant 2s+1$. The quantities $\widehat{k}^{\mu\nu}_{-(l)}(z_-(\lambda),x_B)$ are defined in Eqs. \eqref{eq:kRPMdec}.
    \label{th:theo8}
  \end{theorem}
\end{widetext}

A similar reasoning allows us to state an analogous theorem for the gravitational time delay function at emission. Indeed, the right-hand side of Eq.~\eqref{eq:intdeltEg} involves line integrals along the light path $\widetilde{z}_+(\mu)$ parametrized by $0\leqslant\mu\leqslant1$. After Taylor expanding the light path about the point event $z_+(\mu)$, the right-hand side of Eq.~\eqref{eq:kTay} involves terms such as $k^{\mu\nu}(z_+(\mu))$. Thus, by making use of Eq. \eqref{eq:kPM}, we find
 \begin{equation}
  k^{\mu\nu}(z_+(\mu),G)=\sum_{l=1}^{\infty}k^{\mu\nu}_{(l)}(z_+(\mu))\text{.}
  \label{eq:kPMz+}
\end{equation}

Then the general expansion of $k^{\mu\nu}(\widetilde{z}_+(\mu))$ is obtained after substituting for $\Xi$ from Eq.~\eqref{eq:delRPMN0G}, and for $k^{\mu\nu}$ from Eq.~\eqref{eq:kPMz+}, into \eqref{eq:kTay}. After some algebra, we find
\begin{equation}
  k^{\mu\nu}(\widetilde{z}_+(\mu),N_0,G)=\sum_{l=s}^{\infty}\widehat{k}^{\mu\nu}_{+(l)}(x_A,z_+(\mu))
  \label{eq:kEPM}
\end{equation}
where the quantities $\widehat{k}^{\mu\nu}_{+(l)}(x_A,z_+(\mu))$ are given for $l\geqslant s$ by the following expressions
\begin{subequations}\label{eq:kEPMdec}
\begin{equation}
  \widehat{k}^{\mu\nu}_{+(s)}(x_A,z_+(\mu))=k_{(1)}^{\mu\nu}(z_+(\mu))\text{,}
\end{equation}
and
\begin{align}
  &\widehat{k}^{\mu\nu}_{+(ps)}(x_A,z_+(\mu))=k^{\mu\nu}_{(p)}(z_+(\mu))\nonumber\\
  &+\sum_{m=1}^{p-1}\sum_{n=1}^{ms}\Phi_+^{(ms,n)}(x_A,\mathbf z_+(\mu))\left[\frac{\partial^nk_{(p-m)}^{\mu\nu}}{(\partial x^0)^n}\right]_{z_+(\mu)}\label{eq:kEPMdecl}
\end{align}
for $p\geqslant 2$, and
\begin{align}
  &\widehat{k}^{\mu\nu}_{+(ps+q)}(x_A,z_+(\mu))=\nonumber\\
  &+\sum_{n=1}^{q}\Phi_+^{(q,n)}(x_A,\mathbf z_+(\mu))\left[\frac{\partial^nk_{(p)}^{\mu\nu}}{(\partial x^0)^n}\right]_{z_+(\mu)}\nonumber\\
  &+\sum_{m=1}^{p-1}\sum_{n=1}^{ms+q}\Phi_+^{(ms+q,n)}(x_A,\mathbf z_+(\mu))\left[\frac{\partial^nk_{(p-m)}^{\mu\nu}}{(\partial x^0)^n}\right]_{z_+(\mu)}\label{eq:kEPMdecl}
\end{align}
\end{subequations}\\
for $p\geqslant 1$ and $1\leqslant q\leqslant s-1$ where $p$ and $q$ are determined from $l$ using the relationships in Eq. \eqref{eq:pq}.

By substituting for $k^{\mu\nu}(\widetilde{z}_-(\lambda))$ from Eq. \eqref{eq:kEPM} into \eqref{eq:intdeltEg}, we infer the theorem which follows.

\begin{widetext}
  \begin{theorem}
    In the optical spacetime, within the quasi-Minkowskian path approximation, $\Xi$ admits the series expansion introduced in Theorem \ref{th:theo5}, so the function $\Xi_{\mathrm{g}}$ is given by
    \begin{equation}
      \Xi_{\mathrm{g}}(x_A,\mathbf{x}_B,N_0,G)=\sum_{l=1}^{\infty}\Xi_{\mathrm{g}}^{(l)}(x_A,\mathbf{x}_B)
      \label{eq:delEgPM}
    \end{equation}
    where
    \begin{subequations}\label{eq:intdelEgPM}
    \begin{align}
      \Xi_{\mathrm{g}}^{(1)}(x_A,\mathbf{x}_B)&=\frac{R_{AB}}{2}\int^{1}_{0}\big(k^{00}_{(1)}-2k^{0i}_{(1)}N^i_{AB}+k^{ij}_{(1)}N^i_{AB}N^j_{AB}\big)_{z_+(\mu)}\dd\mu\text{,}\\
      \Xi_{\mathrm{g}}^{(l)}(x_A,\mathbf{x}_B)&=\frac{R_{AB}}{2}\int^{1}_{0}\big(\widehat k^{00}_{+(s+l-1)}-2\widehat k^{0i}_{+(s+l-1)}N^i_{AB}+\widehat k^{ij}_{+(s+l-1)}N^i_{AB}N^j_{AB}\big)_{(x_A,z_+(\mu))}\dd\mu
    \end{align}
    for $2\leqslant l\leqslant s$, and
    \begin{align}
      \Xi_{\mathrm{g}}^{(l)}(x_A,\mathbf{x}_B)&=\frac{R_{AB}}{2}\int^{1}_{0}\Bigg\{\big(\widehat{k}^{00}_{+(s+l-1)}-2\widehat{k}^{0i}_{+(s+l-1)}N^i_{AB}+\widehat{k}^{ij}_{+(s+l-1)}N^i_{AB}N^j_{AB}\big)_{(x_A,z_+(\mu))}\nonumber\\
      &-2\sum_{m=s}^{l-1}\big(\widehat k^{0i}_{+(m)}-\widehat k^{ij}_{+(m)}N_{AB}^j\big)_{(x_A,z_+(\mu))}\left[\frac{\partial\Xi_{\mathrm{g}}^{(l-m)}}{\partial x^i}\right]_{(x_A,\mathbf{z}_+(\mu))}\nonumber\\
      &+\eta^{ij}\sum_{m=1}^{l-s}\left[\frac{\partial\Xi_{\mathrm{g}}^{(m)}}{\partial x^i}\frac{\partial\Xi_{\mathrm{g}}^{(l-s-m+1)}}{\partial x^j}\right]_{(x_A,\mathbf{z}_+(\mu))}\Bigg\}\dd\mu
    \end{align}
    for $s+1\leqslant l\leqslant 2s$, and finally
    \begin{align}
      \Xi_{\mathrm{g}}^{(l)}(x_A,\mathbf{x}_B)&=\frac{R_{AB}}{2}\int^{1}_{0}\Bigg\{\big(\widehat{k}^{00}_{+(s+l-1)}-2\widehat{k}^{0i}_{+(s+l-1)}N^i_{AB}+\widehat{k}^{ij}_{+(s+l-1)}N^i_{AB}N^j_{AB}\big)_{(x_A,z_+(\mu))}\nonumber\\
      &-2\sum_{m=s}^{l-1}\big(\widehat k^{0i}_{+(m)}-\widehat k^{ij}_{+(m)}N_{AB}^j\big)_{(x_A,z_+(\mu))}\left[\frac{\partial\Xi_{\mathrm{g}}^{(l-m)}}{\partial x^i}\right]_{(x_A,\mathbf{z}_+(\mu))}\nonumber\\
      &+\eta^{ij}\sum_{m=1}^{l-s}\left[\frac{\partial\Xi_{\mathrm{g}}^{(m)}}{\partial x^i}\frac{\partial\Xi_{\mathrm{g}}^{(l-s-m+1)}}{\partial x^j}\right]_{(x_A,\mathbf{z}_+(\mu))}\nonumber\\
      &+\sum_{m=s}^{l-s-1}\big(\widehat{k}^{ij}_{+(m)}\big)_{(x_A,z_+(\mu))}\sum_{n=1}^{l-s-m}\left[\frac{\partial\Xi_{\mathrm{g}}^{(n)}}{\partial x^i}\frac{\partial\Xi_{\mathrm{g}}^{(l-s-m-n+1)}}{\partial x^j}\right]_{(x_A,\mathbf{z}_+(\lambda))}\Bigg\}\dd\mu
    \end{align}
    \end{subequations}
    for $l\geqslant 2s+1$. The quantities $\widehat{k}^{\mu\nu}_{+(l)}(x_A,z_+(\mu))$ are defined in Eqs. \eqref{eq:kEPMdec}.
    \label{th:theo9}
  \end{theorem}
\end{widetext}


As a final remark, let us emphasize that Eqs.~\eqref{eq:intdelRrPM1} and \eqref{eq:intdelRgPM1} are independent of the total delay function. Therefore, as mentioned previously, they can be used directly in Eq. \eqref{eq:s} for the determination of $s$.

\subsection{The coupling time delay functions}
\label{sec:detcou}

All the basic ingredients needed for the establishment of the general expansion of the coupling time delay functions, have been introduced in Secs.~\ref{sec:detref} and \ref{sec:detgra}. The general expansions of the refractive and gravitational spacetime perturbations are given in Eqs. \eqref{eq:kapRPM} and \eqref{eq:kRPM}, respectively. Then the expansions of the reception time delay functions can be found in Eqs. \eqref{eq:delRPMN0G}, \eqref{eq:delRrPMN0G}, and \eqref{eq:delRgPMN0G}.

Therefore, by substituting for $\kappa^{\mu\nu}(\widetilde{z}_-(\lambda))$ and $k^{\mu\nu}(\widetilde{z}_-(\lambda))$ from Eqs.~\eqref{eq:kapRPM} and \eqref{eq:kRPM} into \eqref{eq:intdeltRgr}, respectively, we obtain the theorem which follows.

\begin{widetext}
  \begin{theorem}
    In the optical spacetime, within the quasi-Minkowskian path approximation, $\Delta$ admits the series expansion introduced in Theorem \ref{th:theo4}, so the function $\Delta_{\mathrm{gr}}$ is given by
    \begin{equation}
      \Delta_{\mathrm{gr}}(\mathbf{x}_A,x_B,N_0,G)=\sum_{l=1}^{\infty}\Delta_{\mathrm{gr}}^{(l)}(\mathbf{x}_A,x_B)
      \label{eq:delRgrPMN0G}
    \end{equation}
    where
    \begin{subequations}\label{eq:intdelRgrPM}
    \begin{align}
      \Delta_{\mathrm{gr}}^{(l)}\big|_{l\geqslant 1}(\mathbf{x}_A,x_B)&=R_{AB}\int_0^1\Bigg\{\eta^{ij}\sum_{m=1}^{l}\left[\frac{\partial\Delta_{\mathrm{r}}^{(m)}}{\partial x^i}\frac{\partial\Delta_{\mathrm{g}}^{(l-m+1)}}{\partial x^j}\right]_{(\mathbf{z}_-(\lambda),x_B)}\nonumber\\
      &+\sum_{m=1}^{l}\big(\widehat{\kappa}^{0i}_{-(m)}-\widehat{\kappa}^{ij}_{-(m)}N^j_{AB}\big)_{(z_-(\lambda),x_B)}\left[\frac{\partial\Delta_{\mathrm{g}}^{(l-m+1)}}{\partial x^i}\right]_{(\mathbf{z}_-(\lambda),x_B)}\nonumber\\
      &+\sum_{m=s}^{l+s-1}\big(\widehat{k}^{0i}_{-(m)}-\widehat{k}^{ij}_{-(m)}N^j_{AB}\big)_{(z_-(\lambda),x_B)}\left[\frac{\partial\Delta_{\mathrm{r}}^{(l+s-m)}}{\partial x^i}\right]_{(\mathbf{z}_-(\lambda),x_B)}\Bigg\}\dd\lambda
    \end{align}
    for $l\geqslant 1$, and
    \begin{align}
      \Delta_{\mathrm{gr}}^{(l)}\big|_{l\geqslant 2}(\mathbf{x}_A,x_B)&=\Delta_{\mathrm{gr}}^{(l)}\big|_{l\geqslant 1}(\mathbf{x}_A,x_B)\nonumber\\
      &+R_{AB}\int_0^1\Bigg\{\eta^{ij}\sum_{m=1}^{l-1}\left[\frac{\partial\Delta_{\mathrm{r}}^{(m)}}{\partial x^i}\frac{\partial\Delta_{\mathrm{gr}}^{(l-m)}}{\partial x^j}\right]_{(\mathbf{z}_-(\lambda),x_B)}\nonumber\\
      &+\sum_{m=1}^{l-1}\big(\widehat{\kappa}^{0i}_{-(m)}-\widehat{\kappa}^{ij}_{-(m)}N^j_{AB}\big)_{(z_-(\lambda),x_B)}\left[\frac{\partial\Delta_{\mathrm{gr}}^{(l-m)}}{\partial x^i}\right]_{(\mathbf{z}_-(\lambda),x_B)}\nonumber\\
      &+\sum_{m=1}^{l-1}\big(\widehat{\kappa}^{ij}_{-(m)}\big)_{(z_-(\lambda),x_B)}\sum_{n=1}^{l-m}\left[\frac{\partial\Delta_{\mathrm{r}}^{(n)}}{\partial x^i}\frac{\partial\Delta_{\mathrm{g}}^{(l-m-n+1)}}{\partial x^j}\right]_{(\mathbf{z}_-(\lambda),x_B)}\nonumber\\
      &+\frac{1}{2}\sum_{m=s}^{l+s-2}\big(\widehat{k}^{ij}_{-(m)}\big)_{(z_-(\lambda),x_B)}\sum_{n=1}^{l+s-m-1}\left[\frac{\partial\Delta_{\mathrm{r}}^{(n)}}{\partial x^i}\frac{\partial\Delta_{\mathrm{r}}^{(l+s-m-n)}}{\partial x^j}\right]_{(\mathbf{z}_-(\lambda),x_B)}\Bigg\}\dd\lambda
    \end{align}
    for $l\geqslant 2$, and
    \begin{align}
      \Delta_{\mathrm{gr}}^{(l)}\big|_{l\geqslant 3}(\mathbf{x}_A,x_B)&=\Delta_{\mathrm{gr}}^{(l)}\big|_{l\geqslant 2}(\mathbf{x}_A,x_B)\nonumber\\
      &+R_{AB}\int_0^1\Bigg\{\sum_{m=1}^{l-2}\big(\widehat{\kappa}^{ij}_{-(m)}\big)_{(z_-(\lambda),x_B)}\sum_{n=1}^{l-m-1}\left[\frac{\partial\Delta_{\mathrm{r}}^{(n)}}{\partial x^i}\frac{\partial\Delta_{\mathrm{gr}}^{(l-m-n)}}{\partial x^j}\right]_{(\mathbf{z}_-(\lambda),x_B)}\Bigg\}\dd\lambda
    \end{align}
    for $l\geqslant 3$, and
    \begin{align}
      \Delta_{\mathrm{gr}}^{(l)}\big|_{l\geqslant s+1}(\mathbf{x}_A,x_B)&=\Delta_{\mathrm{gr}}^{(l)}\big|_{l\geqslant s}(\mathbf{x}_A,x_B)\nonumber\\
      &+R_{AB}\int_0^1\Bigg\{\eta^{ij}\sum_{m=1}^{l-s}\left[\frac{\partial\Delta_{\mathrm{g}}^{(m)}}{\partial x^i}\frac{\partial\Delta_{\mathrm{gr}}^{(l-s-m+1)}}{\partial x^j}\right]_{(\mathbf{z}_-(\lambda),x_B)}\nonumber\\
      &+\sum_{m=s}^{l-1}\big(\widehat{k}^{0i}_{-(m)}-\widehat{k}^{ij}_{-(m)}N^j_{AB}\big)_{(z_-(\lambda),x_B)}\left[\frac{\partial\Delta_{\mathrm{gr}}^{(l-m)}}{\partial x^i}\right]_{(\mathbf{z}_-(\lambda),x_B)}\nonumber\\
      &+\sum_{m=s}^{l-1}\big(\widehat{k}^{ij}_{-(m)}\big)_{(z_-(\lambda),x_B)}\sum_{n=1}^{l-m}\left[\frac{\partial\Delta_{\mathrm{r}}^{(n)}}{\partial x^i}\frac{\partial\Delta_{\mathrm{g}}^{(l-m-n+1)}}{\partial x^j}\right]_{(\mathbf{z}_-(\lambda),x_B)}\nonumber\\
      &+\frac{1}{2}\sum_{m=1}^{l-s}\big(\widehat{\kappa}^{ij}_{-(m)}\big)_{(z_-(\lambda),x_B)}\sum_{n=1}^{l-s-m+1}\left[\frac{\partial\Delta_{\mathrm{g}}^{(n)}}{\partial x^i}\frac{\partial\Delta_{\mathrm{g}}^{(l-s-m-n+2)}}{\partial x^j}\right]_{(\mathbf{z}_-(\lambda),x_B)}\Bigg\}\dd\lambda
    \end{align}
    for $l\geqslant s+1$, and
    \begin{align}
      \Delta_{\mathrm{gr}}^{(l)}\big|_{l\geqslant s+2}(\mathbf{x}_A,x_B)&=\Delta_{\mathrm{gr}}^{(l)}\big|_{l\geqslant s+1}(\mathbf{x}_A,x_B)\nonumber\\
      &+\frac{R_{AB}}{2}\int_0^1\Bigg\{\eta^{ij}\sum_{m=1}^{l-s-1}\left[\frac{\partial\Delta_{\mathrm{gr}}^{(m)}}{\partial x^i}\frac{\partial\Delta_{\mathrm{gr}}^{(l-s-m)}}{\partial x^j}\right]_{(\mathbf{z}_-(\lambda),x_B)}\nonumber\\
      &+2\sum_{m=s}^{l-2}\big(\widehat{k}^{ij}_{-(m)}\big)_{(z_-(\lambda),x_B)}\sum_{n=1}^{l-m-1}\left[\frac{\partial\Delta_{\mathrm{r}}^{(n)}}{\partial x^i}\frac{\partial\Delta_{\mathrm{gr}}^{(l-m-n)}}{\partial x^j}\right]_{(\mathbf{z}_-(\lambda),x_B)}\nonumber\\
      &+2\sum_{m=1}^{l-s-1}\big(\widehat{\kappa}^{ij}_{-(m)}\big)_{(z_-(\lambda),x_B)}\sum_{n=1}^{l-s-m}\left[\frac{\partial\Delta_{\mathrm{g}}^{(n)}}{\partial x^i}\frac{\partial\Delta_{\mathrm{gr}}^{(l-s-m-n+1)}}{\partial x^j}\right]_{(\mathbf{z}_-(\lambda),x_B)}\Bigg\}\dd\lambda
    \end{align}
    for $l\geqslant s+2$, and
    \begin{align}
      \Delta_{\mathrm{gr}}^{(l)}\big|_{l\geqslant s+3}(\mathbf{x}_A,x_B)&=\Delta_{\mathrm{gr}}^{(l)}\big|_{l\geqslant s+2}(\mathbf{x}_A,x_B)\nonumber\\
      &+\frac{R_{AB}}{2}\int_0^1\Bigg\{\sum_{m=1}^{l-s-2}\big(\widehat{\kappa}^{ij}_{-(m)}\big)_{(z_-(\lambda),x_B)}\sum_{n=1}^{l-s-m-1}\left[\frac{\partial\Delta_{\mathrm{gr}}^{(n)}}{\partial x^i}\frac{\partial\Delta_{\mathrm{gr}}^{(l-s-m-n)}}{\partial x^j}\right]_{(\mathbf{z}_-(\lambda),x_B)}\Bigg\}\dd\lambda
    \end{align}
    for $l\geqslant s+3$, and
    \begin{align}
      \Delta_{\mathrm{gr}}^{(l)}\big|_{l\geqslant 2s+1}(\mathbf{x}_A,x_B)&=\Delta_{\mathrm{gr}}^{(l)}\big|_{l\geqslant 2s}(\mathbf{x}_A,x_B)\nonumber\\
      &+R_{AB}\int_0^1\Bigg\{\sum_{m=s}^{l-s-1}\big(\widehat{k}^{ij}_{-(m)}\big)_{(z_-(\lambda),x_B)}\sum_{n=1}^{l-s-m}\left[\frac{\partial\Delta_{\mathrm{g}}^{(n)}}{\partial x^i}\frac{\partial\Delta_{\mathrm{gr}}^{(l-s-m-n+1)}}{\partial x^j}\right]_{(\mathbf{z}_-(\lambda),x_B)}\Bigg\}\dd\lambda
    \end{align}
    for $l\geqslant 2s+1$, and finally
    \begin{align}
      \Delta_{\mathrm{gr}}^{(l)}\big|_{l\geqslant 2s+2}(\mathbf{x}_A,x_B)&=\Delta_{\mathrm{gr}}^{(l)}\big|_{l\geqslant 2s+1}(\mathbf{x}_A,x_B)\nonumber\\
      &+\frac{R_{AB}}{2}\int_0^1\Bigg\{\sum_{m=s}^{l-s-2}\big(\widehat{k}^{ij}_{-(m)}\big)_{(z_-(\lambda),x_B)}\sum_{n=1}^{l-s-m-1}\left[\frac{\partial\Delta_{\mathrm{gr}}^{(n)}}{\partial x^i}\frac{\partial\Delta_{\mathrm{gr}}^{(l-s-m-n)}}{\partial x^j}\right]_{(\mathbf{z}_-(\lambda),x_B)}\Bigg\}\dd\lambda
    \end{align}
    \end{subequations}
    for $l\geqslant 2s+2$. The quantities $\widehat{k}^{\mu\nu}_{-(l)}(z_-(\lambda),x_B)$ and $\widehat{\kappa}^{\mu\nu}_{-(l)}(z_-(\lambda),x_B)$ are defined in Eqs. \eqref{eq:kRPMdec} and \eqref{eq:kapRPMdec}, respectively.
    \label{th:theo10}
  \end{theorem}
\end{widetext}

Applying the exact same reasoning, the analogous theorem for the coupling time delay function at emission can be stated. Indeed, substituting for $\kappa^{\mu\nu}(\widetilde{z}_+(\mu))$ from Eq. \eqref{eq:kapEPM} and for $k^{\mu\nu}(\widetilde{z}_+(\mu))$ from Eq. \eqref{eq:kEPM} into \eqref{eq:intdeltEgr}, we obtain the theorem which follows.

\begin{widetext}
  \begin{theorem}
    In the optical spacetime, within the quasi-Minkowskian path approximation, $\Xi$ admits the series expansion introduced in Theorem \ref{th:theo5}, so the function $\Xi_{\mathrm{gr}}$ is given by
    \begin{equation}
      \Xi_{\mathrm{gr}}(x_A,\mathbf x_B,N_0,G)=\sum_{l=1}^{\infty}\Xi_{\mathrm{gr}}^{(l)}(x_A,\mathbf x_B)
      \label{eq:delEgrPM}
    \end{equation}
    where
    \begin{subequations}\label{eq:intdelEgrPM}
    \begin{align}
      \Xi_{\mathrm{gr}}^{(l)}\big|_{l\geqslant 1}(x_A,\mathbf x_B)&=R_{AB}\int_0^1\Bigg\{\eta^{ij}\sum_{m=1}^{l}\left[\frac{\partial\Xi_{\mathrm{r}}^{(m)}}{\partial x^i}\frac{\partial\Xi_{\mathrm{g}}^{(l-m+1)}}{\partial x^j}\right]_{(x_A,\mathbf{z}_+(\mu))}\nonumber\\
      &-\sum_{m=1}^{l}\big(\widehat{\kappa}^{0i}_{+(m)}-\widehat{\kappa}^{ij}_{+(m)}N^j_{AB}\big)_{(x_A,z_+(\mu))}\left[\frac{\partial\Xi_{\mathrm{g}}^{(l-m+1)}}{\partial x^i}\right]_{(x_A,\mathbf{z}_+(\mu))}\nonumber\\
      &-\sum_{m=s}^{l+s-1}\big(\widehat{k}^{0i}_{+(m)}-\widehat{k}^{ij}_{+(m)}N^j_{AB}\big)_{(x_A,z_+(\mu))}\left[\frac{\partial\Xi_{\mathrm{r}}^{(l+s-m)}}{\partial x^i}\right]_{(x_A,\mathbf{z}_+(\mu))}\Bigg\}\dd\mu
    \end{align}
    for $l\geqslant 1$, and
    \begin{align}
      \Xi_{\mathrm{gr}}^{(l)}\big|_{l\geqslant 2}(x_A,\mathbf x_B)&=\Xi_{\mathrm{gr}}^{(l)}\big|_{l\geqslant 1}(x_A,\mathbf x_B)\nonumber\\
      &+R_{AB}\int_0^1\Bigg\{\eta^{ij}\sum_{m=1}^{l-1}\left[\frac{\partial\Xi_{\mathrm{r}}^{(m)}}{\partial x^i}\frac{\partial\Xi_{\mathrm{gr}}^{(l-m)}}{\partial x^j}\right]_{(x_A,\mathbf{z}_+(\mu))}\nonumber\\
      &-\sum_{m=1}^{l-1}\big(\widehat{\kappa}^{0i}_{+(m)}-\widehat{\kappa}^{ij}_{+(m)}N^j_{AB}\big)_{(x_A,z_+(\mu))}\left[\frac{\partial\Xi_{\mathrm{gr}}^{(l-m)}}{\partial x^i}\right]_{(x_A,\mathbf{z}_+(\mu))}\nonumber\\
      &+\sum_{m=1}^{l-1}\big(\widehat{\kappa}^{ij}_{+(m)}\big)_{(x_A,z_+(\mu))}\sum_{n=1}^{l-m}\left[\frac{\partial\Xi_{\mathrm{r}}^{(n)}}{\partial x^i}\frac{\partial\Xi_{\mathrm{g}}^{(l-m-n+1)}}{\partial x^j}\right]_{(x_A,\mathbf{z}_+(\mu))}\nonumber\\
      &+\frac{1}{2}\sum_{m=s}^{l+s-2}\big(\widehat{k}^{ij}_{+(m)}\big)_{(x_A,z_+(\mu))}\sum_{n=1}^{l+s-m-1}\left[\frac{\partial\Xi_{\mathrm{r}}^{(n)}}{\partial x^i}\frac{\partial\Xi_{\mathrm{r}}^{(l+s-m-n)}}{\partial x^j}\right]_{(x_A,\mathbf{z}_+(\mu))}\Bigg\}\dd\mu
    \end{align}
    for $l\geqslant 2$, and
    \begin{align}
      \Xi_{\mathrm{gr}}^{(l)}\big|_{l\geqslant 3}(x_A,\mathbf x_B)&=\Xi_{\mathrm{gr}}^{(l)}\big|_{l\geqslant 2}(x_A,\mathbf x_B)\nonumber\\
      &+R_{AB}\int_0^1\Bigg\{\sum_{m=1}^{l-2}\big(\widehat{\kappa}^{ij}_{+(m)}\big)_{(x_A,z_+(\mu))}\sum_{n=1}^{l-m-1}\left[\frac{\partial\Xi_{\mathrm{r}}^{(n)}}{\partial x^i}\frac{\partial\Xi_{\mathrm{gr}}^{(l-m-n)}}{\partial x^j}\right]_{(x_A,\mathbf{z}_+(\mu))}\Bigg\}\dd\mu
    \end{align}
    for $l\geqslant 3$, and
    \begin{align}
      \Xi_{\mathrm{gr}}^{(l)}\big|_{l\geqslant s+1}(x_A,\mathbf x_B)&=\Xi_{\mathrm{gr}}^{(l)}\big|_{l\geqslant s}(x_A,\mathbf x_B)\nonumber\\
      &+R_{AB}\int_0^1\Bigg\{\eta^{ij}\sum_{m=1}^{l-s}\left[\frac{\partial\Xi_{\mathrm{g}}^{(m)}}{\partial x^i}\frac{\partial\Xi_{\mathrm{gr}}^{(l-s-m+1)}}{\partial x^j}\right]_{(x_A,\mathbf{z}_+(\mu))}\nonumber\\
      &-\sum_{m=s}^{l-1}\big(\widehat{k}^{0i}_{+(m)}-\widehat{k}^{ij}_{+(m)}N^j_{AB}\big)_{(x_A,z_+(\mu))}\left[\frac{\partial\Xi_{\mathrm{gr}}^{(l-m)}}{\partial x^i}\right]_{(x_A,\mathbf{z}_+(\mu))}\nonumber\\
      &+\sum_{m=s}^{l-1}\big(\widehat{k}^{ij}_{+(m)}\big)_{(x_A,z_+(\mu))}\sum_{n=1}^{l-m}\left[\frac{\partial\Xi_{\mathrm{r}}^{(n)}}{\partial x^i}\frac{\partial\Xi_{\mathrm{g}}^{(l-m-n+1)}}{\partial x^j}\right]_{(x_A,\mathbf{z}_+(\mu))}\nonumber\\
      &+\frac{1}{2}\sum_{m=1}^{l-s}\big(\widehat{\kappa}^{ij}_{+(m)}\big)_{(x_A,z_+(\mu))}\sum_{n=1}^{l-s-m+1}\left[\frac{\partial\Xi_{\mathrm{g}}^{(n)}}{\partial x^i}\frac{\partial\Xi_{\mathrm{g}}^{(l-s-m-n+2)}}{\partial x^j}\right]_{(x_A,\mathbf{z}_+(\mu))}\Bigg\}\dd\mu
    \end{align}
    for $l\geqslant s+1$, and
    \begin{align}
      \Xi_{\mathrm{gr}}^{(l)}\big|_{l\geqslant s+2}(x_A,\mathbf x_B)&=\Xi_{\mathrm{gr}}^{(l)}\big|_{l\geqslant s+1}(x_A,\mathbf x_B)\nonumber\\
      &+\frac{R_{AB}}{2}\int_0^1\Bigg\{\eta^{ij}\sum_{m=1}^{l-s-1}\left[\frac{\partial\Xi_{\mathrm{gr}}^{(m)}}{\partial x^i}\frac{\partial\Xi_{\mathrm{gr}}^{(l-s-m)}}{\partial x^j}\right]_{(x_A,\mathbf{z}_+(\mu))}\nonumber\\
      &+2\sum_{m=s}^{l-2}\big(\widehat{k}^{ij}_{+(m)}\big)_{(x_A,z_+(\mu))}\sum_{n=1}^{l-m-1}\left[\frac{\partial\Xi_{\mathrm{r}}^{(n)}}{\partial x^i}\frac{\partial\Xi_{\mathrm{gr}}^{(l-m-n)}}{\partial x^j}\right]_{(x_A,\mathbf{z}_+(\mu))}\nonumber\\
      &+2\sum_{m=1}^{l-s-1}\big(\widehat{\kappa}^{ij}_{+(m)}\big)_{(x_A,z_+(\mu))}\sum_{n=1}^{l-s-m}\left[\frac{\partial\Xi_{\mathrm{g}}^{(n)}}{\partial x^i}\frac{\partial\Xi_{\mathrm{gr}}^{(l-s-m-n+1)}}{\partial x^j}\right]_{(x_A,\mathbf{z}_+(\mu))}\Bigg\}\dd\mu
    \end{align}
    for $l\geqslant s+2$, and
    \begin{align}
      \Xi_{\mathrm{gr}}^{(l)}\big|_{l\geqslant s+3}(x_A,\mathbf x_B)&=\Xi_{\mathrm{gr}}^{(l)}\big|_{l\geqslant s+2}(x_A,\mathbf x_B)\nonumber\\
      &+\frac{R_{AB}}{2}\int_0^1\Bigg\{\sum_{m=1}^{l-s-2}\big(\widehat{\kappa}^{ij}_{+(m)}\big)_{(x_A,z_+(\mu))}\sum_{n=1}^{l-s-m-1}\left[\frac{\partial\Xi_{\mathrm{gr}}^{(n)}}{\partial x^i}\frac{\partial\Xi_{\mathrm{gr}}^{(l-s-m-n)}}{\partial x^j}\right]_{(x_A,\mathbf{z}_+(\mu))}\Bigg\}\dd\mu\text{,}
    \end{align}
    for $l\geqslant s+3$, and
    \begin{align}
      \Xi_{\mathrm{gr}}^{(l)}\big|_{l\geqslant 2s+1}(x_A,\mathbf x_B)&=\Xi_{\mathrm{gr}}^{(l)}\big|_{l\geqslant 2s}(x_A,\mathbf x_B)\nonumber\\
      &+R_{AB}\int_0^1\Bigg\{\sum_{m=s}^{l-s-1}\big(\widehat{k}^{ij}_{+(m)}\big)_{(x_A,z_+(\mu))}\sum_{n=1}^{l-s-m}\left[\frac{\partial\Xi_{\mathrm{g}}^{(n)}}{\partial x^i}\frac{\partial\Xi_{\mathrm{gr}}^{(l-s-m-n+1)}}{\partial x^j}\right]_{(x_A,\mathbf{z}_+(\mu))}\Bigg\}\dd\mu
    \end{align}
    for $l\geqslant 2s+1$, and finally
    \begin{align}
      \Xi_{\mathrm{gr}}^{(l)}\big|_{l\geqslant 2s+2}(x_A,\mathbf x_B)&=\Xi_{\mathrm{gr}}^{(l)}\big|_{l\geqslant 2s+1}(x_A,\mathbf x_B)\nonumber\\
      &+\frac{R_{AB}}{2}\int_0^1\Bigg\{\sum_{m=s}^{l-s-2}\big(\widehat{k}^{ij}_{+(m)}\big)_{(x_A,z_+(\mu))}\sum_{n=1}^{l-s-m-1}\left[\frac{\partial\Xi_{\mathrm{gr}}^{(n)}}{\partial x^i}\frac{\partial\Xi_{\mathrm{gr}}^{(l-s-m-n)}}{\partial x^j}\right]_{(x_A,\mathbf{z}_+(\mu))}\Bigg\}\dd\mu
    \end{align}
    \end{subequations}
    for $l\geqslant 2s+2$. The quantities $\widehat{k}^{\mu\nu}_{+(l)}(x_A,z_+(\mu))$ and $\widehat{\kappa}^{\mu\nu}_{+(l)}(x_A,z_+(\mu))$ are defined in Eqs. \eqref{eq:kEPMdec} and \eqref{eq:kapEPMdec}, respectively.
    \label{th:theo11}
  \end{theorem}
\end{widetext}

Finally, from the gravitational, the refractive, and the coupling components, we can now determine the time delay expression up to the $l$th order by applying Theorem \ref{th:theo4}. Then the expressions for the range and the time transfer functions are determined from Eqs. \eqref{eq:RTFdecR} and \eqref{eq:RTFrdef}, respectively.

Let us emphasize that line integrals occurring in Eqs.~\eqref{eq:intdelRrPM}, \eqref{eq:intdelRgPM}, and \eqref{eq:intdelRgrPM} are now zeroth-order null geodesics with parametric equations $x=z_-(\lambda)$. Similarly, Eqs.~\eqref{eq:intdelErPM}, \eqref{eq:intdelEgPM}, and \eqref{eq:intdelEgrPM} are integrated along the zeroth-order null geodesic path with parametric equations $x=z_+(\mu)$. This specificity of the time transfer functions formalism considerably simplifies the integrations and constitutes one of the most important advantages with respect to an explicit resolution of the null geodesic equation \cite{2002PhRvD..66b4045L,2004CQGra..21.4463L,2008CQGra..25n5020T}.

The usefulness of the decomposition performed in Eq.~\eqref{eq:del} becomes really apparent in stationary optical spacetimes. Indeed when the coordinates $(x^\mu)$ are chosen so that the optical spacetime metric does not depend on $x^0$, it is seen that the series expansions in Eqs.~\eqref{eq:kapRPM} and \eqref{eq:kRPM} reduce to
\begin{subequations}\label{eq:kapkRPMsta}
\begin{align}
  \kappa^{\mu\nu}(\widetilde{z}_-(\lambda),N_0,G)&=\sum_{l=1}^{\infty}\widehat{\kappa}^{\mu\nu}_{-(l)}(z_-(\lambda),x_B)\text{,}\label{eq:kapRPMsta}\\
  k^{\mu\nu}(\widetilde{z}_-(\lambda),N_0,G)&=\sum_{l=1}^{\infty}\widehat{k}^{\mu\nu}_{-(l)}(z_-(\lambda),x_B)\label{eq:kRPMsta}
\end{align}
\end{subequations}
where
\begin{subequations}\label{eq:kapkRPMdecsta}
\begin{align}
  \widehat{\kappa}^{\mu\nu}_{(l)}(z_-(\lambda),x_B)&=\kappa^{\mu\nu}_{(l)}(z_-(\lambda))\text{,}\label{eq:kapRPMdecsta}\\
  \widehat{k}^{\mu\nu}_{(ps+q)}(z_-(\lambda),x_B)&=\delta(q)\,k^{\mu\nu}_{(p)}(z_-(\lambda))\text{,}\label{eq:kRPMdecsta}
\end{align}
\end{subequations}
respectively. We recall that $p$ and $q$ are determined from $l$ using Eq. \eqref{eq:pq}. Hence, the different theorems can be solved independently from each other. As a matter of fact, theorems involving gravitational perturbation become independent of $\widehat{k}^{\mu\nu}_{(l)}$ for any $l$ which is not a multiple of $s$.

\section{Application to stationary optical spacetime in geocentric celestial reference system}
\label{sec:app}

Let us now illustrate the method by determining the time transfer function up to the postlinear approximation. We investigate the light-dragging effect experienced by a signal during its propagation inside a flowing media of non-null refractivity. In the GCRS, the effect shows up at the postlinear approximation. In the case where the motion of the Earth's atmosphere is mainly a steady rotation (e.g., in GCRS), we show that the light-dragging effect reduces to a geometrical factor scaling the static atmospheric contribution. During the computation, we never make use of an \emph{a priori} index of refraction profile in order to keep equations as general as possible.

\subsection{Notations and definitions}

We consider that spacetime is covered with some global coordinates $(x^\mu)$. We choose the coordinate system such that the optical metric components are independent of $x^0$. In addition, the coordinate system shall be chosen in such a way that it is convenient to model the outcomes of an experiment taking place in the Earth's close vicinity. Therefore, we consider that $(x^\mu)$ are the GCRS coordinates. We recall that the GCRS is centered in the Earth's center of mass and is nonrotating with respect to distant stars. We suppose that the domain $\mathcal{D}$ defines the spacetime boundaries of the Earth's neutral atmosphere. In that sense, $\mathcal{D}$ draws a timelike tube in spacetime. The Earth's atmosphere is considered spherically symmetric and we suppose that it is filled with a nondispersive fluid dielectric medium whose refractive properties are independent of the component $x^0$, that is to say
\begin{equation}
  n(\mathbf{x})= 1+N(\mathbf{x})\text{.}
\end{equation}
We consider that the atmosphere is still in the reference system rotating with the Earth; thus we assume that the unit 4-velocity vector is given in GCRS by
\begin{equation}
  w^{\mu}=w^0(1,\xi^i)
  \label{eq:4velmed}
\end{equation}
where $\xi^i$ is the coordinate 3-velocity vector of the fluid dielectric medium. Hereafter, we assume that the 3-velocity vector is given by the following expression:
\begin{equation}
  \xi^i(\mathbf x)=\frac{\omega_\oplus}{c}\,e^{ijk}S_\oplus^jx^k
  \label{eq:3velmed}
\end{equation}
where $\omega_\oplus$ is the magnitude of the Earth's angular velocity of rotation and $\mathbf S_\oplus$ is the direction of the spin axis.

Moreover, we consider the case of a one-way transfer, with the transmitter being right outside $\mathcal{D}$ and the receiver being comoving with the fluid dielectrics medium, that is to say at rest in the reference system rotating with the Earth. In order to fix ideas for future discussion, let us assume that the emitter is transmitting from the international space station (ISS) at an altitude of $h\simeq 400\,\mathrm{km}$. Furthermore, let us consider that the emitter is moving along the timelike worldline $\mathcal{C}_A$ with the unit 4-velocity vector $\bm u_{A}$ defined by
\begin{equation}
  u^{\mu}_A=u^0_A(1,\beta^i_A)
  \label{eq:4velA}
\end{equation}
where $\beta^i_A$ is the coordinate 3-velocity vector expressed in GCRS coordinates. Similarly, we assume that the receiver moves along the timelike worldline $\mathcal{C}_B$ with the unit 4-velocity vector $\bm u_{B}$ defined by
\begin{equation}
  u^{\mu}_B=u^0_B(1,\beta^i_B)
  \label{eq:4velB}
\end{equation}
where $\beta^i_B$ is the coordinate 3-velocity vector expressed in GCRS coordinates. For a receiver comoving with the medium, we have
\begin{equation}
  \beta^i_B=\xi^i(\mathbf x_B)\text{.}
\end{equation}

\subsection{Expansion of the delay functions}

The components of the physical spacetime metric expressed in GCRS coordinates are given in \cite{2003AJ....126.2687S} [where the convention for the signature of spacetime is $(-,+,+,+)$ and where the components $G_{\alpha\beta}$ correspond to our $g_{\mu\nu}$]. By keeping terms in $1/c^2$, the first-order gravitational perturbation reads as follows
\begin{subequations}\label{eq:gGCRS}
  \begin{equation}
    h_{00}^{(1)}=-\frac{2U}{c^2}\text{,} \qquad h_{0i}^{(1)}=0\text{,} \qquad h_{ij}^{(1)}=-\frac{2U}{c^2}\delta_{ij}
    \label{eq:gGCRScov}
  \end{equation}
where the contravariant components are determined from Eq. \eqref{eq:kPMord}
  \begin{equation}
    k^{00}_{(1)}=\frac{2U}{c^2}\text{,} \qquad k^{0i}_{(1)}=0\text{,} \qquad k^{ij}_{(1)}=\frac{2U}{c^2}\delta_{ij}\text{.}
    \label{eq:gGCRScon}
  \end{equation}
\end{subequations}

In these expressions, we restrict $U$ to the monopole term of the Newtonian gravitational potential of the Earth, that is
\begin{equation}
  U(\mathbf x)=\frac{Gm_\oplus}{|\mathbf x|}
  \label{eq:U}
\end{equation}
where $m_\oplus$ is the mass of the Earth. In that respect, at the level of the surface of the Earth, we find
\begin{equation}
  |k^{\mu\nu}_{(1)}|_{\mathrm{max}}\propto \frac{U(R_\oplus)}{c^2}\sim10^{-10}
  \label{eq:ordk}
\end{equation}
where $R_\oplus$ denotes the Earth's equatorial radii.

Then according to \cite{doi1010292010JD015214}, at the sea level an average parcel of air possesses a refractivity $N(R_\oplus)\simeq 3\times10^{-4}$, so we consider $N_0=N(R_\oplus)\sim10^{-4}$. Additionally, at the Earth's surface, the 3-velocity of the refractive medium expressed in GCRS coordinates is $|\xi^i(R_\oplus)|_{\mathrm{max}}\propto\omega_\oplus R_\oplus/c\sim 10^{-6}$. Consequently, we can expand the refractive perturbation in terms of the refractivity at the Earth's surface and in the approximation of small velocities. Therefore, it can be seen that the first-order term of the refractive perturbation is given by [see Eqs.~\eqref{eq:kapGCRS1}]
\begin{equation}
  |\kappa^{\mu\nu}_{(1)}|_{\mathrm{max}}\propto N_0\sim 10^{-4}\text{.}
  \label{eq:ordkap}
\end{equation}

At the same time, a typical measurement profile for the neutral atmosphere using global positioning system meteorology occultations data \cite{1999AnGeo..17..122S} starts at $\ell\simeq100\,\mathrm{km}$, so that $\ell/h\simeq 0.4$. For observations at lower elevation than the zenith direction, we can roughly take $\ell/R_{AB}\sim0.1$. Then if we consider that the light path is sufficiently small so that the metric components vary slowly during the integration, we can get a rough estimation of $s$ by making use of Eqs. \eqref{eq:kkap0}. We quickly infer that $s$ must satisfy the zeroth-order following relation
\begin{equation}
  \frac{\ell}{R_{AB}}|\kappa^{\mu\nu}|_{\mathrm{max}}\sim (|k^{\mu\nu}|_{\mathrm{max}})^{1/s}\text{.}
  \label{eq:relmag}
\end{equation}
Inserting numerical values, we deduce $s=2$. These results can be double checked by inserting the first-order expressions of the gravitational and refractive delays [see Eqs.~\eqref{eq:delShap} and \eqref{eq:delr1GCRS}] into Eq. \eqref{eq:s}.

In this application, we exclude third-order terms and beyond, that is to say, all terms of the order of $\varepsilon^3$ with $\varepsilon\sim\ell/R_{AB}|\kappa^{\mu\nu}|_{\mathrm{max}}\sim 10^{-5}$. The meaning is that a postlinear expression of the range transfer function neglects terms of the order of $\varepsilon^3R_{AB}$. Therefore, the coupling terms which are of third order are neglected too.

A look at Eqs.~\eqref{eq:Gorcon} and \eqref{eq:ordkap} allows one to infer that the time component of the 4-velocity vector of the fluid dielectric medium must be known up to $10^{-5}$ in order to account for all second-order terms. Considering that the 4-velocity of the medium must be a unit vector for the spacetime metric $g_{\mu\nu}$, we have the relation
\begin{equation}
  w^0=\left(g_{00}+2g_{0i}\xi^i+g_{ij}\xi^i\xi^j\right)^{-1/2}\text{.}
  \label{eq:u0}
\end{equation}
Therefore, to sufficient accuracy, we can safely consider for the rest of the application that $w^0=1$.

Hence, we end up with the following contravariant components for the refractive perturbation:
\begin{subequations}\label{eq:kapGCRS}
  \begin{equation}
    \kappa^{00}_{(1)}=2N\text{,} \qquad \kappa^{0i}_{(1)}=0\text{,} \qquad \kappa^{ij}_{(1)}=0\text{,}
    \label{eq:kapGCRS1}
  \end{equation}
  with the second order
  \begin{equation}
    \kappa^{00}_{(2)}=N^2\text{,} \qquad \kappa_{(2)}^{0i}=2N\xi^i\text{,} \qquad \kappa_{(2)}^{ij}=0\text{.}
    \label{eq:kapGCRS2}
  \end{equation}
\end{subequations}
Let us note that the cross component is non-null at the postlinear approximation. It represents the light-dragging effect due to the motion of the fluid dielectric medium in GCRS coordinates.

Additionally, let us mention that the optical spacetime is stationary as seen from Eqs.~\eqref{eq:gGCRScon} and \eqref{eq:kapGCRS}. In that respect, the emission or the reception time transfer functions become identical. As a consequence, the distinction between emission and reception functions is not relevant anymore meaning that the time component at emission or reception is no longer an independent variable \cite{2008CQGra..25n5020T}. Hence, $\widehat \kappa_{(l)}^{\mu\nu}$ and $\widehat k_{(l)}^{\mu\nu}$ are now given by Eqs. \eqref{eq:kapkRPMsta} and \eqref{eq:kapkRPMdecsta} which are independent of the total time delay. Therefore, the refractive and the gravitational delays may be solved independently from each other.

A straightforward application of Theorem \ref{th:theo4} assuming $s=2$, allows us to infer the expansion scheme of the total time delay function
\begin{subequations}\label{eq:ord2}
  \begin{align}
    \Delta^{(1)}(\mathbf{x}_A,\mathbf x_B)&=\Delta^{(1)}_{\mathrm r}(\mathbf{x}_A,\mathbf x_B)\text{,}\\
    \Delta^{(2)}(\mathbf{x}_A,\mathbf x_B)&=\Delta^{(2)}_{\mathrm r}(\mathbf{x}_A,\mathbf x_B)+\Delta^{(1)}_{\mathrm g}(\mathbf{x}_A,\mathbf x_B)\text{.}
  \end{align}
\end{subequations}
Thus, we deduce the fact that the different contributions in Eq.~\eqref{eq:delR} are given by 
\begin{subequations}
\begin{align}
  \Delta_{\mathrm g}(\mathbf{x}_A,\mathbf x_B)&=\Delta^{(1)}_{\mathrm g}(\mathbf{x}_A,\mathbf x_B)\text{,}\\
  \Delta_{\mathrm r}(\mathbf{x}_A,\mathbf x_B)&=\Delta^{(1)}_{\mathrm r}(\mathbf{x}_A,\mathbf x_B)+\Delta^{(2)}_{\mathrm r}(\mathbf{x}_A,\mathbf x_B)\text{.}
\end{align}
\end{subequations}
Then Theorems \ref{th:theo6} and \ref{th:theo8} together with Eqs. \eqref{eq:kapGCRS} and \eqref{eq:gGCRScon} allow us to determine the refractive and the gravitational contributions up to the appropriate order.

\subsection{Time transfer function and Doppler}

Using the fact that spacetime is stationary, we first focus on the gravitational time delay. By making use of Theorem \ref{th:theo8}, we soon arrive at the well known formula
\begin{equation}
  \Delta_{\mathrm g}(\mathbf{x}_A,\mathbf x_B)=\frac{2R_{AB}}{c^2}\int_0^1U(\mathbf z_-(\lambda))\,\dd\lambda\text{,}
\end{equation}
which leads after integration to the Shapiro delay \cite{1964PhRvL..13..789S}
\begin{equation}
  \Delta_{\mathrm g}(\mathbf{x}_A,\mathbf x_B)=\frac{2Gm_\oplus}{c^2}\mathrm{ln}\left(\frac{r_A+r_B+R_{AB}}{r_A+r_B-R_{AB}}\right)\text{.}\label{eq:delShap}
\end{equation}
We introduced the notations $r_{A/B}=|\mathbf{x}_{A/B}|$.

The first-order refractive contribution is derived from Theorem \ref{th:theo6} and is given by
\begin{equation}
  \Delta^{(1)}_{\mathrm{r}}(\mathbf{x}_A,\mathbf x_B)=R_{AB}\int_0^1N(\mathbf z_-(\lambda))\,\dd\lambda\text{.}
  \label{eq:delr1GCRS}
\end{equation}
We find almost similar expressions for the atmospheric delay in \cite{inproceedingsMendes,2002GeoRL..29.1414M,2004GeoRL..3114602M,2010ITN....36....1P} (commonly, when applied to the Earth's neutral atmosphere, the refractivity is defined within a factor of $10^6$ and is separated into hydrostatic and a nonhydrostatic components). The first main difference stands in the path of integration in Eq. \eqref{eq:delr1GCRS} which is performed along the Euclidean line between the emitter and the receiver even for nonzenithal observations. Instead, in the literature [cf., e.g., Eqs.~(2-3) of \cite{2002GeoRL..29.1414M}], the atmospheric delay is usually computed at zenith, and then mapping functions are used to convert the zenithal delay into a delay in the line-of-sight direction as discussed in \cite{2002GeoRL..29.1414M}. The other difference stands in the upper limit of integration. However, considering that the refractive region is bounded to the domain $\mathcal{D}$ of spacetime, the integration out of $\mathcal{D}$ does not contribute to the final results. In that respect, the difference in the upper integration limit is only superficial.

The first-order refractive delay \eqref{eq:delr1GCRS} is the well known excess path delay due to the change of the phase velocity experienced by the signal during the crossing of the dielectric medium. The geometric delay due to the refractive bending of the ray arises at the postlinear order as we shall see in the next paragraph.

According to Theorem \ref{th:theo6}, the second order is given by
\begin{align}
  \Delta^{(2)}_{\mathrm r}(\mathbf{x}_A,\mathbf x_B)&=\frac{R_{AB}}{2}\int_0^1\Big\{\big(N^2-4N\xi^iN_{AB}^i\big)_{\mathbf z_-(\lambda)}\nonumber\\
  &-\big[\partial_i\Delta^{(1)}_{\mathrm r}\partial_i\Delta^{(1)}_{\mathrm r}\big]_{(\mathbf z_-(\lambda),\mathbf x_B)}\Big\}\dd\lambda\text{.}\label{eq:delr2GCRS}
\end{align}
The term $\partial_i\Delta^{(1)}_{\mathrm r}$ is computed by differentiating Eq. \eqref{eq:delr1GCRS} with respect to $x_A^i$, that is to say
\begin{align}
  \big[\partial_i\Delta^{(1)}_{\mathrm r}\big]_{(\mathbf x,\mathbf x_B)}&=-\frac{(\mathbf x_B-\mathbf x)^i}{|\mathbf x_B-\mathbf x|}\int_0^1N(\mathbf y_-(\mu,\mathbf x))\,\dd\mu\nonumber\\
  &+|\mathbf x_B-\mathbf x|\int_0^1\mu\big[\partial_i N\big]_{\mathbf y_-(\mu,\mathbf x)}\dd\mu\text{.}\label{eq:DPdel}
\end{align}
We have introduced
\begin{equation}
  \mathbf y_-(\mu,\mathbf x)=(1-\mu)\,\mathbf x_B+\mu\,\mathbf x\text{,}
\end{equation}
which reduces to
\begin{equation}
  \mathbf y_-(\mu,\mathbf z_-(\lambda))=\mathbf z_-(\mu\lambda)
  \label{eq:ym}
\end{equation}
when $\mathbf x=\mathbf z_-(\lambda)$,

We can rearrange Eq. \eqref{eq:delr2GCRS} by first noticing that the light-dragging contribution can be further simplified. Indeed, after making use of Eq. \eqref{eq:3velmed}, it may be seen that
\begin{equation}
  \big(\xi^iN_{AB}^i\big)_{\mathbf{z}_-(\lambda)}\equiv\xi^i(\mathbf x_B)N_{AB}^i
\end{equation}
which is obviously independent of $\lambda$. Then by substituting for $\partial_i\Delta^{(1)}_{\mathrm r}$ from Eq. \eqref{eq:DPdel} into \eqref{eq:delr2GCRS} while accounting for Eq. \eqref{eq:ym} and \eqref{eq:z-i}, one can apply the following change of variables $\mu'=\mu\lambda$, and by integrating by parts the double integrals, one infers the postlinear refractive order
\begin{align}
  \Delta^{(2)}_{\mathrm{r}}(\mathbf{x}_A,\mathbf x_B)&=\Delta^{(2)}_{\mathrm{r,exc}}(\mathbf{x}_A,\mathbf x_B)+\Delta^{(2)}_{\mathrm{r,geo}}(\mathbf{x}_A,\mathbf x_B)\nonumber\\
  &+\Delta^{(2)}_{\mathrm{r,drag}}(\mathbf{x}_A,\mathbf x_B)\label{eq:delr2}
\end{align}
where
\begin{subequations}\label{eq:delr2GCRSbis}
\begin{align}
  \Delta^{(2)}_{\mathrm{r,exc}}(\mathbf{x}_A,\mathbf x_B)&=\frac{R_{AB}}{2}\int_0^1N^2(\mathbf z_-(\lambda))(1+\mathrm{ln}\lambda)\,\dd\lambda\text{,}\label{eq:delr2GCRSexc}\\
  \Delta^{(2)}_{\mathrm{r,geo}}(\mathbf{x}_A,\mathbf x_B)&=\frac{R_{AB}}{2}\int_0^1\Big\{\lambda R_{AB}^2\big[\partial_i N\partial_i N\big]_{\mathbf z_-(\lambda)}\nonumber\\
  &-2R_{AB}N_{AB}^i \big[N\partial_i N\big]_{\mathbf z_-(\lambda)}\Big\}\lambda\mathrm{ln}\lambda\,\dd\lambda\text{,}\label{eq:delr2GCRSgeo}
\end{align}
and
\begin{equation}
  \Delta^{(2)}_{\mathrm{r,drag}}(\mathbf{x}_A,\mathbf x_B)=D(\mathbf x_A,\mathbf x_B)\,\Delta^{(1)}_{\mathrm r}(\mathbf{x}_A,\mathbf x_B)\text{,}
  \label{eq:delr2GCRSdrag}
\end{equation}
\end{subequations}
with $D(\mathbf x_A,\mathbf x_B)$ being given by
\begin{equation}
  D(\mathbf x_A,\mathbf x_B)=-2\xi^i(\mathbf x_B)N_{AB}^i\text{.}
  \label{eq:Kdrag}
\end{equation}

We have separated the postlinear approximation of the refractive time delay function into three components. The first one, namely Eq.~\eqref{eq:delr2GCRSexc}, is the second-order correction to the excess path delay \eqref{eq:delr1GCRS}. The second component, that is, Eq.~\eqref{eq:delr2GCRSgeo}, is the geometric delay which accounts for the bending of the ray. These two components together with Eq.~\eqref{eq:delr1GCRS} constitute the static part of the refractive time delay
\begin{align}
  \Delta_{\mathrm{r,stat}}(\mathbf x_A,\mathbf x_B)&=\Delta_{\mathrm r}^{(1)}(\mathbf x_A,\mathbf x_B)+\Delta_{\mathrm{r,exc}}^{(2)}(\mathbf x_A,\mathbf x_B)\nonumber\\
  &+\Delta_{\mathrm{r,geo}}^{(2)}(\mathbf x_A,\mathbf x_B)\text{,}\label{eq:delrstat}
\end{align}
namely the refractive part of delay that would be measured or modeled in a frame comoving with the media. Instead, the last term in Eq. \eqref{eq:delr2}, namely Eq.~\eqref{eq:delr2GCRSdrag}, is the delay due to the dragging of light caused by the motion of the dielectric medium. In that respect, $D(\mathbf x_A,\mathbf x_B)$ is referred to as the light-dragging factor.

Interestingly, one might see from Eq.~\eqref{eq:delr2GCRSdrag} that the light-dragging contribution can be expressed as a geometric factor scaling the first order of the static refractive part. This fact is not a specificity of the postlinear approximation but must hold true for higher order terms too. Indeed, it results from the really specific form of the refractive components $\kappa^{0i}$ which can always be written as
\begin{equation}
  \kappa^{0i}=\kappa^{00}\xi^i\text{.}
\end{equation}
Therefore, because the scalar product $\xi^iN_{AB}^i$ is independent of the path of integration for a steady rotating atmosphere, the integration of $\kappa^{0i}N_{AB}^i$ reduces to
\begin{equation}
  \xi^i(\mathbf x_B)N_{AB}^i\int_0^1(\kappa^{00})_{\mathbf z_-(\lambda)}\dd\lambda
\end{equation}
where the integrated term corresponds to the static part of the refraction.

Solving the line integrals in Eqs. \eqref{eq:delr1GCRS}, \eqref{eq:delr2GCRSexc}, and \eqref{eq:delr2GCRSgeo} for a realistic index of refraction is not an easy task. Moreover it is beyond the scope of this paper which aims at introducing a recursive method allowing one to determine the integral form of the time transfer functions up to any order in optical spacetime. For this reason, we address the effective resolution of the line integrals to future work. Hereafter, we derive the range and the time transfer function at the postlinear approximation.

From Eqs.~\eqref{eq:RTFdecR}, and by making use of Eqs.~\eqref{eq:ord2}, \eqref{eq:delr2}, and \eqref{eq:delr2GCRSdrag}, we find
\begin{align}
  \mathcal{R}(\mathbf x_A,\mathbf x_B)&=R_{AB}+\Delta_{\mathrm{g}}(\mathbf{x}_A,\mathbf x_B)\nonumber\\
  &+C(\mathbf x_A,\mathbf x_B)\,\Delta^{(1)}_{\mathrm{r}}(\mathbf{x}_A,\mathbf x_B)\nonumber\\
  &+\Delta^{(2)}_{\mathrm{r,exc}}(\mathbf{x}_A,\mathbf x_B)+\Delta^{(2)}_{\mathrm{r,geo}}(\mathbf{x}_A,\mathbf x_B)\label{eq:RTFpl}
\end{align}
where we have introduced the factor $C(\mathbf x_A,\mathbf x_B)$ being defined such that
\begin{equation}
  C(\mathbf x_A,\mathbf x_B)=1+D(\mathbf x_A,\mathbf x_B)\text{.}
  \label{eq:Ccoeff}
\end{equation}

According to previous discussions, we can rewrite Eq.~\eqref{eq:RTFpl}, within the same accuracy, such as
\begin{align}
  \mathcal{R}(\mathbf x_A,\mathbf x_B)&=R_{AB}+\Delta_{\mathrm{g}}(\mathbf{x}_A,\mathbf x_B)\nonumber\\
  &+C(\mathbf x_A,\mathbf x_B)\,\Delta_{\mathrm{r,stat}}(\mathbf{x}_A,\mathbf x_B)\text{.}\label{eq:RTFplD}
\end{align}
The time transfer function can be directly obtained by making use of Eq. \eqref{eq:RTFrdef}
\begin{align}
  \mathcal{T}(\mathbf x_A,\mathbf x_B)&=\frac{1}{c}\Big[R_{AB}+\Delta_{\mathrm{g}}(\mathbf{x}_A,\mathbf x_B)\nonumber\\
  &+C(\mathbf x_A,\mathbf x_B)\,\Delta_{\mathrm{r,stat}}(\mathbf{x}_A,\mathbf x_B)\Big]\label{eq:TTFpl}
\end{align}
where we recall that $\Delta_{\mathrm{r,stat}}$ is given in Eq.~\eqref{eq:delrstat}.

Let us emphasize how simple result \eqref{eq:TTFpl} is. As a matter of fact, the light-dragging effect is enclosed into a geometrical factor scaling the static part of the refractive delay. In addition, to derive \eqref{eq:TTFpl} we never made use of an \emph{a priori} refractive profile; we only supposed a stationary rotating optical medium. In comparison, a derivation of the light-dragging effect using perturbation equations applied to geometrical optics \cite{2019AeA...624A..41B} requires heavier calculations (where the integration must be performed along an hyperbolic path) highlighting the advantage of using the covariant formalism developed so far. Indeed, in a covariant theory, the light-dragging contribution is naturally taken into account through the cross components of Gordon's metric.

From the range or the time transfer functions, we can derive the expression of the frequency transfer within the postlinear approximation as well. After inserting Eq.~\eqref{eq:RTFplD} into \eqref{eq:qTF}, we deduce
\begin{subequations}\label{eq:qAqB}
\begin{align}
  q_A&=1-\beta^i_AN_{AB}^i+\beta^i_A\frac{\partial\Delta_{\mathrm{g}}}{\partial x_A^i}+\widehat\beta^i_A\frac{\partial\Delta_{\mathrm{r,stat}}}{\partial x_A^i}\nonumber\\
  &+\beta^i_A\frac{\partial D}{\partial x_A^i}\Delta_{\mathrm{r,stat}}\text{,}
\end{align}
and
\begin{align}
  q_B&=1-\beta^i_BN_{AB}^i-\beta^i_B\frac{\partial\Delta_{\mathrm{g}}}{\partial x_B^i}-\widehat\beta^i_B\frac{\partial\Delta_{\mathrm{r,stat}}}{\partial x_B^i}\nonumber\\
  &-\beta^i_B\frac{\partial D}{\partial x_B^i}\Delta_{\mathrm{r,stat}}
\end{align}
\end{subequations}
where we have introduced two artificial ``dragging'' coordinate velocities defined by
\begin{equation}
  \widehat\beta^i_{A/B}=C(\mathbf x_A,\mathbf x_B)\,\beta^i_{A/B}\text{.}
  \label{eq:velD}
\end{equation}

Most of the time, while modeling range and Doppler observables in GCRS coordinates, the factor $C$ is arbitrarily fixed to $C=1$ (i.e., vanishing of the light-dragging factor). In the next, we investigate the resulting consequences by discussing orders of magnitude and variabilities due to the light-dragging contribution in the expressions of the time and the frequency transfers.

\subsection{Light-dragging magnitude and variability}

In GCRS coordinates, the velocity of the fluid medium at $\mathbf x_B$ is given by Eq. \eqref{eq:3velmed}, that is
\begin{equation}
  \xi^i(\mathbf x_B)=\frac{\omega_\oplus r_B}{c}\,e^{ijk}S_\oplus^jn_B^k
  \label{eq:3velmedN}
\end{equation}
where $\mathbf n_B=\mathbf x_B/r_B$. For a ground-based receiver, we have $r_B=R_\oplus$ and the light-dragging factor becomes
\begin{equation}
  D(\mathbf x_A,\mathbf x_B)=-\frac{2\omega_\oplus R_\oplus}{c}\,(\mathbf S_\oplus\times\mathbf n_B)\cdot\mathbf N_{AB}\text{.}
  \label{eq:KdragGCRS}
\end{equation}
Thus, the maximum value of $D$ is about
\begin{equation}
  \frac{2\omega_\oplus R_\oplus}{c}\simeq 3.099\times 10^{-6}\text{.}
  \label{eq:Knum}
\end{equation}

A typical value of the static refractive delay in the zenith direction is approximately $2.5\,\mathrm m$ and can reach $15\,\mathrm m$ for sn elevation angle of $10\degres$ \cite{1997JGR...10220489C,doi1010292006JB004834}. Therefore, the light-dragging contribution to the time transfer is expected to remain lower that $0.05\,\mathrm{mm}$ in GCRS coordinates. However, for experiments whose data are mainly analyzed in the Barycentric Celestial Reference System (BCRS), the velocity of the media possesses an orbital component which is of the order of $30\,\mathrm{km}\cdot\mathrm{s}^{-1}$. Thus, the maximum value of $D$ becomes of the order of $2\times10^{-4}$, and the dragging contribution can reach $3\,\mathrm{mm}$ in BCRS coordinates.

Experiments such like satellite or lunar laser ranging are currently operating at the millimeter and centimeter levels of precision on range measurements \cite{1998A&AS..130..235S,2013RPPh...76g6901M,2017A&A...602A..90C}. Therefore, the light-dragging effect is just below the threshold of visibility on both experiments. However, as may be inferred from Eq.~\eqref{eq:KdragGCRS}, the effect is mainly suppressed in the case of a round-trip light path. In other words, it might play a significant role only for one-way and three-way configurations.

From Eq. \eqref{eq:DPdel}, considering a slowly varying refractivity, we can infer that
\begin{equation*}
  \frac{\partial\Delta_{\mathrm{r,stat}}}{\partial x_A^i}\sim\frac{\ell}{R_{AB}}N_0N_{AB}^i\sim10^{-5}N_{AB}^i\text{,}
\end{equation*}
hence
\begin{equation}
  \beta_A^i\frac{\partial\Delta_{\mathrm{r,stat}}}{\partial x_A^i}\sim10^{-5}\,(\beta_A^iN_{AB}^i)\text{.}
\end{equation}
Therefore, for a one-way frequency transfer experiment, the static atmospheric contribution relative to the classical effect $(\beta_A^iN_{AB}^i)$, represents roughly 1 part in $10^{5}$.

Then the contribution due to the dragging velocity in Eqs. \eqref{eq:qAqB} is approximately given by
\begin{equation}
  \widehat{\beta}_A^i\frac{\partial\Delta_{\mathrm{r,stat}}}{\partial x_A^i}\sim10^{-5}\,(\widehat{\beta}_A^iN_{AB}^i)\text{.}
\end{equation}
Making use of Eqs. \eqref{eq:velD} and \eqref{eq:Knum}, one infers that, in GCRS coordinates, the light-dragging contribution (term proportional to $D$) represents 1 part in $10^6$ and 1 part in $10^{11}$ relative to the static atmospheric effect and to the classical effect, respectively. If we take a look at orders of magnitude in BCRS coordinates, the light-dragging contribution relative to the the static atmospheric effect reaches 1 part in $10^4$ and 1 part in $10^9$ relative to the classical effect. Therefore, for typical spacecraft's velocities of $10^{-5}$ and $10^{-4}$ in GCRS and BCRS coordinates, respectively, one infers that the effect of the light-dragging contribution produces a fractional frequency change of the order of 1 part in $10^{16}$ in GCRS coordinates and one part in $10^{13}$ in BCRS coordinates. For one-way radio links, these fractional frequency changes translate into radio signal frequencies at the level of $1\,\mu\mathrm{Hz}$ for $\mathrm{X}/\mathrm{Ka}$-bands and $0.1\,\mu\mathrm{Hz}$ for $\mathrm{S}$-bands in GCRS coordinates. In BCRS coordinates, the frequencies of the radio signal due to the dragging of light should arise at $1\mathrm{mHz}$ for $\mathrm{X}/\mathrm{Ka}$-bands and $0.1\,\mathrm{mHz}$ for $\mathrm{S}$-bands. The correspondence in term of velocity precision in the Doppler is at the level of $0.01\,\mu\mathrm{m}\cdot\mathrm{s}^{-1}$ and $10\,\mu\mathrm{m}\cdot\mathrm{s}^{-1}$ in GCRS and BCRS coordinates, respectively.

Past and future space missions such as Cassini \cite{2003Natur.425..374B,Kliore2004,2007IJMPD..16.2117I}, BepiColombo \cite{2002PhRvD..66h2001M,2009AcAau..65..666I}, or JUICE \cite{Grasset20131} have reached or will reach the level of $1\,\mu\mathrm{m}\cdot\mathrm{s}^{-1}$ for the Doppler. Therefore, the light-dragging effect is clearly at the threshold of visibility in Doppler observables and should be modeled in data reduction software in the near future.

In order to understand what could be the signature of an unaccounted light-dragging effect, let us now focus on the computation of the time variability of $D(\mathbf x_A,\mathbf x_B)$. For a ground-based instrument, the spatial coordinates expressed in an Earth centered frame are given by $\mathbf x_B=(R_\oplus,\phi_B,\lambda_B)$, where $\phi_B$ is the latitude and $\lambda_B$ the longitude of the instrument on the surface of the Earth. The variable part in Eq. \eqref{eq:KdragGCRS} is better understood if we introduce $(a,e,\iota,\varOmega,\omega,f)$ denoting the set of Keplerian elements of the emitter. In GCRS coordinates the direction $\mathbf n_A$ of the emitter is given for instance in Eq.~(3.42) of \cite{2014gravbookP}. Then the expression of the light-dragging factor reads as follows:
\begin{align}
  &D=\frac{2\omega_\oplus R_\oplus}{c}\frac{a(1-e^2)}{R_{AB}}\frac{\cos\phi_B}{(1+e\cos f)}\nonumber\\*
  &\times\Big\{\sin\varOmega\big[I_-\cos(F_++P_+)+I_+\cos(F_-+P_-)\big]\nonumber\\*
  &-\cos\varOmega\big[I_-\sin(F_++P_+)+I_+\sin(F_-+P_-)\big]\Big\}
\end{align}
where we have set
\begin{subequations}
\begin{equation}
  I_\pm=\frac{1}{2}\pm\frac{\cos\iota}{2}\text{,}
\end{equation}
and
\begin{equation}
  F_\pm=f\pm \omega_\oplus t\text{,} \qquad P_\pm=\omega\pm\lambda_B\text{.}
\end{equation}
\end{subequations}

Considering a quasicircular orbit ($e\ll 1$), we have $r_A=a+\mathcal{O}(e)$ and
\begin{equation}
  f=n (t-t_0)+\mathcal{O}(e)
\end{equation}
where $t_0$ is the time of perigee passage and where $n$ is the mean motion being given by Kepler's third law
\begin{equation}
  n=\sqrt{\frac{Gm_\oplus}{a^3}}\,\text{.}
\end{equation}
Therefore, the magnitude of $D$ oscillates with frequencies $n\pm\omega_\oplus$ around zero and $10^{-4}$ (maximum amplitude of the orbital barycentric velocity) in GCRS and BCRS coordinates, respectively. The peak to peak amplitude is of the order of $10^{-6}$ in both reference systems. In the limit case where $\lim_{a\to\infty}n=0$, the same magnitudes oscillate at diurnal frequency.

Consequently, while modeling the time and frequency transfers using Eqs. \eqref{eq:TTFpl} and \eqref{eq:qAqB} in GCRS or BCRS coordinates, the fact of imposing $C=1$ (or equivalently $D=0$) leads to an unaccounted contribution which may lead to systematic errors for instance in the estimations of the spacecraft velocity [considering Eq. \eqref{eq:velD}] or in the receiver coordinates (considering that diurnal signatures mainly concern ground-based stations). This last example could be particularly relevant for ground-based techniques operating within the international Earth rotation and reference system service (IERS) for which an error in the estimation of the station coordinates can result in a bias in the determination of the ITRF.

\section{Conclusion}

This paper generalizes the algorithmic approach introduced in \cite{2008CQGra..25n5020T} by making the time transfer functions formalism applicable in optical spacetime. The main results stand in the Theorems~\ref{th:theo4}--\ref{th:theo11} which allow one to determine the integral form of the time transfer functions up to any order. The great benefit of using the time transfer functions formalism relies on the fact that all integrals in Theorems \ref{th:theo6}--\ref{th:theo11} are line integrals taken along the zeroth-order null geodesic path between the emitter and the receiver, independently of the order being considered.

In optical spacetime, the method requires us to know the order of magnitudes of both the gravitational and the refractive perturbations. Then one can deduce the integer parameter $s$ from Eq. \eqref{eq:s} and use Theorems \ref{th:theo4}--\ref{th:theo5} in order to determine the general expansion of the total time delay functions. The different components are the gravitational, the refractive, and the coupling contributions. Each of them is determined recursively making use of Theorems~\ref{th:theo7}--\ref{th:theo11}. We emphasize that these theorems have been derived assuming (i) a post-Minkowskian expansion and (ii) a general expansion in terms of an arbitrary refractivity $N_0$. Both choices are motivated by the quasi-Minkowskian path regime which is assumed throughout the paper.

We have illustrated the method by determining the integral form of the time transfer function up to the postlinear approximation. We have considered the case of a one-way transfer between a low orbit emitter and a receiving station on the Earth's surface. We have shown that the time and frequency transfers are both impacted by the light-dragging effect due to the motion of the atmosphere, as seen from a frame which is not comoving with the flowing optical media. With respect to other methods (e.g., \cite{2019AeA...624A..41B}), we have highlighted the great advantage of the covariant formalism developed in this paper which naturally takes into account the effect of the dragging of light. In addition, we have shown that the light-dragging contribution is independent of the refractive profile which is considered. At the end of the day, the dragging component reduces to a geometrical factor which scales the static part of the atmospheric time delay (where the term ``static'' refers to the delay which would be measured in a frame comoving with the refractive medium). Concerning the frequency transfer, we have shown that the light-dragging contribution scales the coordinate velocities of both the emitter and the receiver resulting in the introduction of artificial dragging coordinate velocities. Finally, we have discussed the necessity, in the near future, for taking into account the dragging of light in data reduction software modeling the time and frequency transfers within GCRS or BCRS coordinates.

\section*{Acknowledgments}

The author is grateful to the University of Bologna and to the Italian Space Agency (ASI) for financial support through Agreement No. 2013-056-RO in the context of EAS's JUICE mission. The author is also thankful to A. Hees, P. Teyssandier, and C. Le Poncin-Lafitte from SYRTE in Observatoire de Paris for interesting discussions and valuable comments about a preliminary version of the manuscript.

\appendix

\section{General expansion of $\gamma_{\mu\nu}$}
\label{sec:genexp}

The covariant components of $\gamma_{\mu\nu}$ are determined from the inverse conditions which lead to the following implicit expression
\begin{equation}
  \gamma_{\mu\nu}=-g_{\mu\alpha}g_{\beta\nu}\kappa^{\alpha\beta}-g_{\mu\alpha}\kappa^{\alpha\beta}\gamma_{\beta\nu}\text{.}
  \label{eq:gmrec}
\end{equation}
Usually, assuming that $\gamma_{\mu\nu}=f(n)w^{\mu}w^{\nu}$ with $f(n)$ being a sought function of the index of refraction and using Eq.~\eqref{eq:Gorcon}, we infer Eq.~\eqref{eq:Gorcov}. However, the situation slightly changes if we expand the contravariant components $\kappa^{\mu\nu}$ as is done in Eq. \eqref{eq:kapPM}.

At the same time, we have assumed that the physical spacetime metric, which is given in Eq. \eqref{eq:gmcov}, satisfies a post-Minkowskian expansion [see Eq. \eqref{eq:hPM}]. Thus, considering that the refractive components are the dominant order according to Eq. \eqref{eq:s} for $s\in\mathbb{N}_{>0}$, we deduce that the covariant components $\gamma_{\mu\nu}$ satisfy the following expansion:
\begin{equation}
  \gamma_{\mu\nu}(x,N_0,G)=\sum_{l=1}^{\infty}\gamma_{\mu\nu}^{(l)}(x)
  \label{eq:gamPM}
\end{equation}
where the quantities $\gamma_{\mu\nu}^{(l)}$ can be recursively determined from Eq. \eqref{eq:gmrec}, that is
\begin{subequations}\label{eq:kapPMord}
\begin{align}
  \gamma_{\mu\nu}^{(1)}&=-\eta_{\mu\alpha}\eta_{\beta\nu}\kappa^{\alpha\beta}_{(1)}\text{,}\label{eq:kapPMord1}\\
  \gamma_{\mu\nu}^{(q)}&=-\eta_{\mu\alpha}\eta_{\beta\nu}\kappa^{\alpha\beta}_{(q)}-\eta_{\mu\alpha}\sum_{m=1}^{q-1}\kappa^{\alpha\beta}_{(m)}\gamma_{\beta\nu}^{(q-m)}\label{eq:kapPMord2}
\end{align}
for $2\leqslant q\leqslant s$, and
\begin{align}
  \gamma_{\mu\nu}^{(s+1)}&=-\eta_{\mu\alpha}\eta_{\beta\nu}\kappa^{\alpha\beta}_{(s+1)}-2\eta_{\mu\alpha}\kappa^{\alpha\beta}_{(1)}h_{\beta\nu}^{(1)}\nonumber\\
  &-\eta_{\mu\alpha}\sum_{m=1}^{s}\kappa^{\alpha\beta}_{(m)}\gamma_{\beta\nu}^{(s-m+1)}\text{,}\label{eq:kapPMord3}\\
  \gamma_{\mu\nu}^{(s+q)}&=-\eta_{\mu\alpha}\eta_{\beta\nu}\kappa^{\alpha\beta}_{(s+q)}-2\eta_{\mu\alpha}\kappa^{\alpha\beta}_{(q)}h_{\beta\nu}^{(1)}\nonumber\\
  &-\eta_{\mu\alpha}\sum_{m=1}^{s+q-1}\kappa^{\alpha\beta}_{(m)}\gamma_{\beta\nu}^{(s-m+q)}\nonumber\\
  &-h_{\mu\alpha}^{(1)}\sum_{m=1}^{q-1}\kappa^{\alpha\beta}_{(m)}\gamma_{\beta\nu}^{(q-m)}\label{eq:kapPMord4}
\end{align}
for $2\leqslant q\leqslant s$, and
\begin{align}
  \gamma_{\mu\nu}^{(2s+1)}&=-\eta_{\mu\alpha}\eta_{\beta\nu}\kappa^{\alpha\beta}_{(2s+1)}-2\eta_{\mu\alpha}\kappa^{\alpha\beta}_{(s+1)}h_{\beta\nu}^{(1)}\nonumber\\
  &-2\eta_{\mu\alpha}\kappa^{\alpha\beta}_{(1)}h_{\beta\nu}^{(2)}-h_{\mu\alpha}^{(1)}\kappa^{\alpha\beta}_{(1)}h^{(1)}_{\beta\nu}\nonumber\\
  &-\eta_{\mu\alpha}\sum_{m=1}^{2s}\kappa^{\alpha\beta}_{(m)}\gamma_{\beta\nu}^{(2s-m+1)}\nonumber\\
  &-h_{\mu\alpha}^{(1)}\sum_{m=1}^{s}\kappa^{\alpha\beta}_{(m)}\gamma_{\beta\nu}^{(s-m+1)}\text{,}\label{eq:kapPMord7}
\end{align}
and
\begin{align}
  \gamma_{\mu\nu}^{(2s+q)}&=-\eta_{\mu\alpha}\eta_{\beta\nu}\kappa^{\alpha\beta}_{(2s+q)}-2\eta_{\mu\alpha}\kappa^{\alpha\beta}_{(q)}h_{\beta\nu}^{(2)}\nonumber\\
  &-2\eta_{\mu\alpha}\kappa^{\alpha\beta}_{(s+q)}h_{\beta\nu}^{(1)}-h_{\mu\alpha}^{(1)}\kappa^{\alpha\beta}_{(q)}h^{(1)}_{\beta\nu}\nonumber\\
  &-\eta_{\mu\alpha}\sum_{m=1}^{2s+q-1}\kappa^{\alpha\beta}_{(m)}\gamma_{\beta\nu}^{(2s+q-m)}\nonumber\\
  &-h_{\mu\alpha}^{(1)}\sum_{m=1}^{s+q-1}\kappa^{\alpha\beta}_{(m)}\gamma_{\beta\nu}^{(s+q-m)}\nonumber\\
  &-h_{\mu\alpha}^{(2)}\sum_{m=1}^{q-1}\kappa^{\alpha\beta}_{(m)}\gamma_{\beta\nu}^{(q-m)}\label{eq:kapPMord8}
\end{align}
for $2\leqslant q\leqslant s$, and 
\begin{align}
  \gamma_{\mu\nu}^{(ps+1)}&=-\eta_{\mu\alpha}\eta_{\beta\nu}\kappa^{\alpha\beta}_{(ps+1)}-2\eta_{\mu\alpha}\sum_{m=0}^{p-1}\kappa^{\alpha\beta}_{(ms+1)}h_{\beta\nu}^{(p-m)}\nonumber\\
  &-\sum_{m=1}^{p-1}h_{\mu\alpha}^{(p-m)}\sum_{n=0}^{m-1}\kappa^{\alpha\beta}_{(ns+1)}h^{(m-n)}_{\beta\nu}\nonumber\\
  &-\eta_{\mu\alpha}\sum_{m=1}^{ps}\kappa^{\alpha\beta}_{(m)}\gamma_{\beta\nu}^{(ps+1-m)}\nonumber\\
  &-\sum_{m=1}^{p-1}h_{\mu\alpha}^{(p-m)}\sum_{n=1}^{ms}\kappa^{\alpha\beta}_{(n)}\gamma_{\beta\nu}^{(ms+1-n)}\label{eq:kapPMord9}
\end{align}
for $p\geqslant 3$, and finally
\begin{align}
  \gamma_{\mu\nu}^{(ps+q)}&=-\eta_{\mu\alpha}\eta_{\beta\nu}\kappa^{\alpha\beta}_{(ps+q)}-2\eta_{\mu\alpha}\sum_{m=0}^{p-1}\kappa^{\alpha\beta}_{(ms+q)}h_{\beta\nu}^{(p-m)}\nonumber\\
  &-\sum_{m=1}^{p-1}h_{\mu\alpha}^{(p-m)}\sum_{n=0}^{m-1}\kappa^{\alpha\beta}_{(ns+q)}h^{(m-n)}_{\beta\nu}\nonumber\\
  &-\eta_{\mu\alpha}\sum_{m=1}^{ps+q-1}\kappa^{\alpha\beta}_{(m)}\gamma_{\beta\nu}^{(ps+q-m)}\nonumber\\
  &-\sum_{m=0}^{p-1}h_{\mu\alpha}^{(p-m)}\sum_{n=1}^{ms+q-1}\kappa^{\alpha\beta}_{(n)}\gamma_{\beta\nu}^{(ms+q-n)}\label{eq:kapPMord10}
\end{align}
\end{subequations}
for $p\geqslant 3$ and $2\leqslant q\leqslant s$, where $p$ and $q$ are determined from $l$ using Eqs. \eqref{eq:pq}.

\bibliographystyle{apsrev4-1}
\bibliography{TTF_opt}

\end{document}